\documentclass[aps,pre,preprint,superscriptaddress]{revtex4-2}

\usepackage{graphicx}
\usepackage{float}
\usepackage{bm}
\usepackage[export]{adjustbox}
\usepackage{color}

\newcommand	{\ovEf}	{\overline{E_f}}

\newcommand	{\EMJ}	{E_{\text{MJ}}}
\newcommand	{\rv}		{\mathbf{r}}
\newcommand  {\lv}           {\bm{\lambda}}

\begin{document}



\title{Folding lattice proteins confined on minimal grids\\ using a quantum-inspired encoding}



\author{Anders Irb\"ack}
\email[]{anders.irback@cec.lu.se}
\affiliation{Computational Science for Health and Environment (COSHE), Centre for Environmental and Climate Science, Lund University, 223 62 Lund, Sweden}


\author{Lucas Knuthson}
\affiliation{Computational Science for Health and Environment (COSHE), Centre for Environmental and Climate Science, Lund University, 223 62 Lund, Sweden}

\author{Sandipan Mohanty}
\affiliation{J\"ulich Supercomputing Centre, Forschungszentrum J\"ulich, D-52425 J\"ulich, Germany}



\begin{abstract}

Steric clashes pose a challenge when exploring dense protein systems
using conventional explicit-chain methods. A minimal example is a single
lattice protein confined on a minimal grid, with no free sites.
Finding its minimum energy is a hard optimization problem, with
similarities to scheduling problems. It can be recast
as a quadratic unconstrained binary optimization (QUBO) problem amenable
to classical and quantum approaches. We show that this problem in its QUBO form
 can be swiftly and consistently solved for chain length 48, using either
 classical simulated annealing or hybrid quantum-classical annealing on a D-Wave system.
 In fact, the latter computations required about 10 seconds. We also test linear and quadratic
 programming methods, which work well for a lattice gas but struggle
 with chain constraints. All methods are benchmarked against exact
 results obtained from exhaustive structure enumeration, at
 a high computational cost.

\end{abstract}


\maketitle


\newpage


\section{Introduction\label{sec:intro}}

Recent years have seen big advances in predicting the native structure of
folded proteins, through the development of evolutionary-informed machine learning
techniques~\cite{Jumper:21}. Still, other biomolecular problems, such
as protein aggregation and phase separation, remain challenging. These problems
often concern dense systems. Exploring dense biomolecular systems by
conventional simulations is computationally demanding. A major obstacle is steric clashes,
which obstruct chain motions. This problem is present even in simplified
models where the molecules are represented as self-avoiding chains on a lattice.

A simple yet extreme example is that of a single lattice protein
confined on a maximally compact grid, where no site is left unoccupied.
Studies of maximally compact lattice models can provide insights into the sequence-structure
relationship of proteins~\cite{Lau:89,Sali:94,Socci:94,Klimov:95,Li:96}. However,
for maximally compact systems, standard Monte Carlo methods are impractical,
due to difficulties in proposing new feasible states. To find the
minimum-energy structure of a given sequence, most studies have instead relied on
exhaustive enumerations of all states, which is an option for short chains.

Another approach to the challenge posed by steric clashes is to switch
from the original explicit-chain representation to an expanded binary
representation, and thus recast the energy minimization problem as a
quadratic unconstrained binary optimization (QUBO) task. In the binary system,
chain structure is enforced through soft constraints, implemented as
energy penalties~\cite{Perdomo-Ortiz:12,Robert:21,Irback:22}.
Mapping hard discrete optimization problems onto binary systems
has a long history~\cite{Hopfield:85,Peterson:88}, and has seen a revival with
the advent of quantum optimization~\cite{Kadowaki:98,Fahri:01,Fahri:14,Hadfield:19}
and various classical Ising machines~\cite{Mohseni:22}. The approach has been
applied to lattice protein~\cite{Perdomo-Ortiz:12,Robert:21,Irback:22}
and homopolymer~\cite{Micheletti:21,Slongo:23} problems, as well as to somewhat
related scheduling tasks~\cite{Lagerholm:97,Venturelli:16,Stollenwerk:20} which
involve restrictions similar to steric clashes.

In this paper, we examine the utility of a binary representation in the determination of
minimum-energy structures of maximally compact lattice proteins.
We explore and compare three methods for optimizing the binary system.
The first two are classical simulated annealing (SA)~\cite{Kirkpatrick:83}
and hybrid quantum-classical annealing as offerred by D-Wave (HA)~\cite{McGeoch:20b},
both of which solve a QUBO variant of the problem with soft constraints.
In addition, we perform calculations with the Gurobi optimizer (GO)~\cite{gurobi}, which works with hard
constraints and combines linear/quadratic programming with heuristics. The results obtained
with these three methods are evaluated against exact data, which we generate using much more
time-consuming exhaustive enumeration techniques.

We consider lattice proteins in 3D, with a 20-letter amino acid alphabet
and pairwise interaction energies given by the statistical Miyazawa-Jernigan
potential~\cite{Miyazawa:85,Miyazawa:96}. Specifically, we search for minimum-energy structures
for six different amino acid sequences with length $N=48$ (Table~\ref{tab:seq}) on a $4\times 4\times 3$ lattice. This was the largest
problem size for which exact results could be generated with the resources at our disposal.

\begin{table}[b]
  \centering
  \caption{Amino acid sequences studied, in one-letter code. All six sequences fold to one of two topologies
  called A and B with low and high complexity, respectively~\cite{Faisca:06}.
  The minimum energy, $\EMJ^{\min}$, is indicated.
  \label{tab:seq}}
  \vspace{12pt}
  \begin{tabular}{cccc}
    \hline
    \# &  Topology & Sequence & $\EMJ^{\min}$ \\
    \hline
    \vspace{-4pt}
    1 & A & \texttt{FRTRPLNHDF\ YNYKIWEPFK\ PADFPKAWDR\ MLDHVWDSMA\ SWGHQHCS}	& $-25.85$\\
    \vspace{-4pt}
    2 & A & \texttt{CDLPPFTYRH\ HGNDFWKNYE\ MIKHWDLWRD\ MFRAFWSDPV\ KASPHQAS}	& $-25.92$\\
    \vspace{-4pt}
    3 & A & \texttt{FRTPWVSHQF\ YAYKLMEHFK\ WGDFCRNMDK\ WIDSLPDRWN\ PAPHDHAS}	& $-26.09$\\
    \vspace{-4pt}
    4 & B & \texttt{KDKIHFRMNY\ GYPAWDAQSV\ KDLTCPRDWH\ FPHMRDPSHN\ WELAFFWS}	& $-25.87$\\
    \vspace{-4pt}
    5 & B & \texttt{ENDVTMDMDP\ SPCLFRIHNL\ PRAHSFDRFG\ WHQFDKYHYK\ WKWAWAPS}	& $-26.15$\\
    \vspace{-4pt}
    6 & B & \texttt{EHDAQLDFDW\ SRWTWHGRNS\ YHAPAMYRWP\ VHDMDKPNPK\ FKIFFLCS} 	& $-26.24$\\
    \hline
  \end{tabular}
\end{table}

The results presented below show that the two QUBO-based approaches, HA and SA, are able to quickly and
reliably solve the energy minimization task for these sequences. With HA, it took the
order of 10\,s to find the solutions, whereas the exhaustive enumeration code took
the order of 10\,h to run.


\section{Methods\label{sec:methods}}

\subsection{Biophysical  model\label{sec:methods_mj}}

We consider a coarse-grained protein model, where
the protein is represented as a self-avoiding chain of $N$ amino acids, or beads,
on a lattice. The interaction potential is pairwise additive and stipulates that two amino acids interact
only if they are nearest neighbors on the lattice but not along the chain. Let $\rv_i$ and $a_i$
denote, respectively, the position and type of  amino acid $i$ ($i=1,\ldots,N$).
The interaction energy, $\EMJ$, can then be written as
\begin{equation}
\EMJ=-\sum_{1\le i<j-1\le N}C(a_i,a_j)\Delta(\rv_i,\rv_j),
\label{eq:explicit-chain-E}\end{equation}
where $\Delta(\rv_i,\rv_j)=1$ if $\rv_i$ and $\rv_j$ are nearest neighbors on the lattice, and $\Delta(\rv_i,\rv_j)=0$
otherwise.

In prior work, we explored the application of HA to fold~\cite{Irback:22} and design~\cite{Irback:24} lattice
proteins in a minimal model of this form, namely the two-letter HP model on
a 2D square lattice~\cite{Lau:89}. In that model, the amino acids are either hydrophobic (H) or polar (P).
The contact energy is given by $C(a_i,a_j)=1$ if $a_i=a_j=$\,H, and $C(a_i,a_j)=0$ otherwise.

In the present paper, we use a 20-letter amino acid alphabet, a 3D cubic lattice, and contact energies
given by the statistically derived Miyazawa-Jernigan interaction matrix~(Table VI of Ref.~\cite{Miyazawa:85}).
Specifically, we minimize $\EMJ$ [Eq.~(\ref{eq:explicit-chain-E})] for six different 48-amino acid sequences (Table~\ref{tab:seq})
confined on a minimal $4\times4\times3$ grid, without any free sites. Despite the restriction to maximally compact
states, there are $\approx$\,$1.3\times10^{11}$ possible conformations available for these chains~\cite{Pande:94,Kloczkowski:97}.
These same six sequences have been studied previously by Fa\'\i sca and Plaxco,
using an unrestricted grid~\cite{Faisca:06}. The sequences were designed to fold into
one of two distinct topologies, A and B, with, respectively, low and high contact order~\cite{Faisca:06}.
The contact order of a given conformation is defined as the average sequence separation between
amino acids that are in contact, and provides a measure of the complexity of the structure. It has been shown to be
inversely correlated with the rate at which folding occurs for small single-domain real proteins~\cite{Plaxco:98}.
In line with this, in kinetic explicit-chain Monte Carlo simulations, Fa\'\i sca and Plaxco found that the sequences with topology A (1--3)
indeed folded faster than those with topology B (4--6)~\cite{Faisca:06}.


\subsection{QUBO representation\label{sec:methods_qubo}}

We wish to determine maximally compact lattice protein structures with minimum energy without having to
resort to exhaustive enumeration of an exponentially large number of states. Unfortunately, conventional
Monte Carlo methods for polymers are impractical for exploring chain structures on a
minimal grid, since steric clashes render the acceptance probability very low, if not zero.

A possible way forward
is to take a QUBO approach, where the chain is mapped onto a binary system and chain structure is enforced
through penalty energies. For this, one may use a turn-based mapping, in which the bits encode the directions
of the links along the chain~\cite{Perdomo-Ortiz:12,Robert:21}. However, it then becomes very cumbersome to
implement non-local interactions along the chain, such as chain self-avoidance.

Instead, we use a field-like
representation with bits at all lattice sites~\cite{Irback:22}, which greatly facilitates the implementation of
non-local interactions and thereby enables the study of longer chains. This representation uses more bits than
what is needed to describe the very shape of a chain, which may seem wasteful. 
However, the total bit count is not worse than with turn-based methods~\cite{Linn:24}, since 
no additional auxiliary bits are needed in order to have a quadratic energy function, as required on D-Wave's systems. Another
field-like binary encoding of lattice proteins, though without the possibility of freezing the sequence, was recently developed
by mapping the system to a lattice gauge theory~\cite{Panizza:25}.     

Our field-like representation uses bits $b_{i,n}$ that indicate whether amino acid $i$ is located on
lattice site $n$ ($b_{i,n}=1$) or not ($b_{i,n}=0$). The energy function is given by~\cite{Irback:22}
\begin{equation}
	E=\EMJ+\lambda_1 E_1+ \lambda_2 E_2\ + \lambda_3 E_3,
	\label{eq:E}
\end{equation}
where $\EMJ$ is the interaction potential [Eq.~(\ref{eq:explicit-chain-E})] and the remaining three terms
are penalty energies whose strengths are controlled by the Lagrange parameters $\lambda_i$.
All terms have closed-form expressions valid for an arbitrary lattice and chain length $N$, which, in brief, are as follows~\cite{Irback:22}.

\begin{itemize}
\item  In a valid chain conformation,
the interaction potential $\EMJ$ [Eq.~(\ref{eq:explicit-chain-E})] can be written as
\begin{equation}
\label{eq:EMJ}
\EMJ=-\sum_{| i-j |>1}C(a_i,a_j)\sum_{\langle n,m\rangle}b_{i,n}b_{j,m},
\end{equation}
where the second sum runs over all nearest-neighbor pairs $n,m$.

\item The penalty energy $E_1$ is given by
\begin{equation}
E_1= \sum_i\left(\sum_n b_{i,n}-1\right)^2,
\label{eq:E1}
\end{equation}
and serves to ensure that each amino acid is located at exactly one lattice site.

\item The penalty energy $E_2$ enforces chain self-avoidance and can be written as
\begin{equation}
E_2=\frac{1}{2} \sum_n \sum_{i\ne j} b_{i,n} b_{j,n}.
\label{eq:E2}
\end{equation}

\item The penalty energy $E_3$ is responsible for chain connectivity and given by
\begin{equation}
E_3= \sum_{1\le i<N}\sum_n b_{i,n}
\sum_{||m-n||>1}b_{i+1,m},
\label{eq:E3}
\end{equation}
where $||m-n||$ denotes the Eucledian distance between $m$ and $n$
(unit lattice spacing).
\end{itemize}

We minimize the QUBO energy $E$ [Eq.~(\ref{eq:E})], using HA or SA, in order
to minimize $\EMJ$ [Eq.~(\ref{eq:explicit-chain-E})] over chain conformations.
For this to work, the Lagrange parameters $\lambda_i$ must be above some thresholds.
Beyond these thresholds, it turns out that the success rates of QUBO HA and QUBO SA
are robust to small changes in the $\lambda_i$ parameters~\cite{Irback:22} (see also Appendix~\ref{sec:app_lagrange}).
In particular, this robustness enables us to use the same parameters,
$\lv=(\lambda_1,\lambda_2,\lambda_3)=(1.5,2.0,2.0)$,
for all six sequences studied (Table~\ref{tab:seq}).


\subsection{Hybrid quantum-classical annealing (HA) \label{sec:methods_ha}}

D-Wave offers access to solvers based solely on quantum annealing as well as hybrid quantum-classical solvers.
The latter approach uses classical solvers while sending suitable subproblems as queries to the quantum processing
unit (QPU). The solutions to the subproblems serve to guide the classical solvers \cite{McGeoch:20b}. The goal is to speed up
the solution of challenging QUBO problems by queries to the QPU. The hybrid approach makes it possible to
tackle problems with many thousands of fully connected variables, which is far beyond what currently can be dealt
with using only the QPU.

Unless specified by the user, the D-Wave hybrid annealer selects the run time based on the problem size.
This default value is also the shortest run time that can be used. This run time may or may not be
sufficiently long for satisfactory results~\cite{Irback:22}.
For each of the six sequences in Table~\ref{tab:seq}, we did HA computations for one or more run times.
For each run time, we generated a set of 100 runs.  Starting with the default value (6\,s for all six systems),
we increased the run time in steps of 2\,s until all 100 runs returned the correct solution, which
gave maximum required run times varying between 6\,s and 22\,s.


\subsection{Simulated annealing (SA)\label{sec:methods_sa}}

A standard method for energy minimization in complex systems is Monte Carlo-based SA~\cite{Kirkpatrick:83}.
Here, one performs constant-temperature updates of the system at hand, as defined by the Boltzmann distribution
$\propto$\,$e^{-\beta E}$, at a sequence of decreasing temperatures ($\beta$ is inverse temperature).

We conducted SA computations for the six sequences in Table~\ref{tab:seq}, starting from random initial bit configurations.
All runs spanned the same set of 25 geometrically distributed temperatures, given by $\beta_i=1.05^i$ ($i=1,\ldots,25$).
The updates were single-bit flips, controlled by a Metropolis acceptance criterion.

To assess the run-time dependence of the results, we generated runs comprising between 1,000 and 80,000 sweeps per temperature,
where one sweep corresponds to, on average, one attempted update per bit variable. One sweep required
approximately 1.1\,ms on an Intel Core i9-13900K processor. For each choice of sequence and run length, we
performed a set of 100 runs, using different random number seeds.


\subsection{Gurobi optimizer (GO)\label{sec:methods_go}}

Besides HA and SA, we also tested using the GO solver~\cite{gurobi}, which
combines simplex~\cite{Dantzig:63}, branch-and-bound~\cite{Morrison:16} and cutting-plane~\cite{Marchand:02} methods with heuristics
to optimize linear or quadratic objective functions under linear or quadratic constraints. 
Here, the penalty energies in Eq.~(\ref{eq:E}) were replaced by hard constraints, whereas
the binary representation of the systems remained the same as in our QUBO computations.
In recent work on protein sequence optimization, the GO solver performed
better than QUBO methods~\cite{Panizza:24}.

We ran the GO program locally under an academic license. For each sequence (Table~\ref{tab:seq}),
we varied the run time from 400\,s to 1200\,s in steps of 200\,s. For each combination of sequence and run time,
we performed 10 runs using five threads each on an Intel Core i9-13900K. Adding more threads
did not further improve the performance.


\subsection{Enumeration of all structures \label{sec:methods_enumeration}}

For our 48-amino acid sequences on a $4\times 4\times 3$ grid, the number of distinct structures
unrelated by symmetry is 134,131,827,475~\cite{Pande:94,Kloczkowski:97}.

To be able to compare our HA, SA and GO results with exact data, we wrote and ran a straight forward
C++ program implementing a depth first graph search algorithm to generate these structures (see Appendix~\ref{sec:app_enumeration}). Using this data,
we first confirmed the previously reported~\cite{Faisca:06} minimum-energy structures.
For each sequence, we then determined the density of states, calculated as a
function of $\EMJ$. In addition, to compare the energy landscapes, we computed for each sequence
the structural similarity with the minimum-energy state for the 100 lowest-energy states.

The calculations were done on the JUSUF cluster at the Jülich Supercomputing Centre,
and required a total of 17 hours of runtime on 512 processor cores (AMD EPYC 7742) for all six sequences.


\section{Results\label{sec:results}}

Conventional sampling methods struggle with dense biomolecular systems,
where chain motions are hampered by steric clashes. Here, we consider
a minimal yet challenging example of such a system, namely a single lattice
protein confined on a minimal grid. Exploring the state space of this system
with standard explicit-chain Monte Carlo methods is impractical, due to steric
clashes. To overcome this issue, we adopt a binary representation (Sec.~\ref{sec:methods_qubo}),
which yields an expanded state space and also opens it up for quantum optimization. Specifically, we use
this representation to minimize the energy of the lattice protein, as given by a pairwise
Miyazawa-Jernigan potential ($\EMJ$; Sec.~\ref{sec:methods_mj}).
We test three methods for solving this binary optimization task.
The first two, HA  (Sec.~\ref{sec:methods_ha}) and SA (Sec.~\ref{sec:methods_sa}), are QUBO methods,
where chain structure is enforced through penalty energies ($E_1$, $E_2$, $E_3$). The third
method, GO (Sec.~\ref{sec:methods_go}), works with hard constraints. As a test bed, we use
a set of six 48-amino acid sequences (Table~\ref{tab:seq}), all of which we study on
a $4\times 4\times 3$ grid.

\begin{figure}[t]
\centering
   \includegraphics[width=8cm]{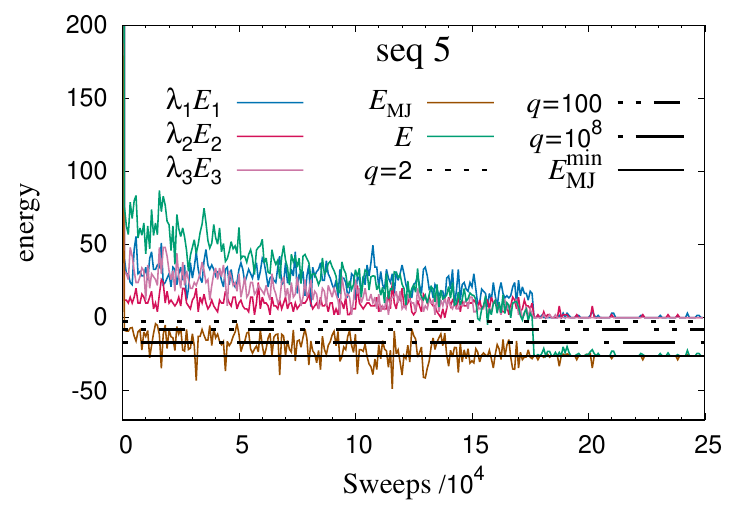}
\caption{Run-time evolution of the interaction potential $\EMJ$ and the
penalty energies $E_1$, $E_2$ and  $E_3$ [Eqs.~(\ref{eq:EMJ}--\ref{eq:E3})]
in an SA run for sequence 5 (Table~\ref{tab:seq}), using $\lv=(1.5,2.0,2.0)$, 25 temperatures
and 10,000 sweeps at each temperature.
The horizontal lines indicate data on the distribution of $\EMJ$ over valid chain structures,
as obtained by exhaustive enumerations (Sec.~\ref{sec:methods_enumeration}).
The lowest line represents the minimum $\EMJ$, $\EMJ^{\min}$. The other three lines represent
the lowest $q$-quantiles for $q=2$ (median), $q=100$ and $q=10^8$, respectively.
\label{fig:runtime}}
\end{figure}

Figure~\ref{fig:runtime} shows the run-time evolution of the interaction potential $\EMJ$ and the
penalty energies $E_1$, $E_2$ and $E_3$ in an SA run for sequence 5 (Table~\ref{tab:seq}), using
25 temperatures and $10^4$ Monte Carlo sweeps (Sec.~\ref{sec:methods_sa}) at each temperature.
The acceptance rate for the Metropolis update is modest, varying from $\approx$\,0.01 at the higher temperatures
to $\approx$\,$5 \times 10^{-5}$ at the lowest temperature. Nevertheless, the system is capable of evolving
toward more chain-like structures with lower penalty energies, and toward the known minimum $\EMJ$ value.
The correct minimum-energy conformation is indeed found.

For reference, Fig.~\ref{fig:runtime} allows shows the lowest $q$-quantiles of the $\EMJ$ probability distribution
for $q=2$ (median), $q=100$ and $q=10^8$, as obtained by exhaustive enumeration of all structures (Sec.~\ref{sec:methods_enumeration}).
The system visits the correct solution for the first time while at the 18th temperature. After this point, the system is
occasionally found in unphysical states, but it stays close to the correct solution and well below the lowest
$10^8$-quantile in terms of $\EMJ$.


\subsection{Comparing HA, SA and GO \label{sec:results_timeevo}}

We wish to assess the ability of the HA, SA and GO methods to locate
minimum-$\EMJ$ chain conformations, and its run-time dependence,
for the six sequences in Table~\ref{tab:seq}. For a given sequence and run time, we generate a set of
10 (GO) or 100 (HA and SA) independent runs. We denote the final energy $E$ [Eq.~(\ref{eq:E})] in a
given run by $E_f$, and the average of $E_f$ over the independent runs by $\ovEf$.
The success rate is defined as the fraction of the runs that return the correct solution, $\EMJ^{\min}$. Determining
a $100\%$ success rate is equivalent to determining that $\ovEf=\EMJ^{\min}$.

\begin{figure}
\centering
   \includegraphics[width=5.4cm]{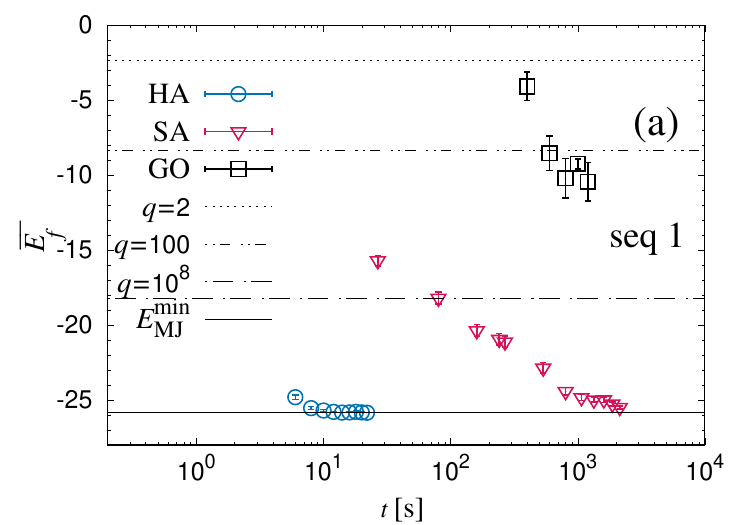}
   \includegraphics[width=5.4cm]{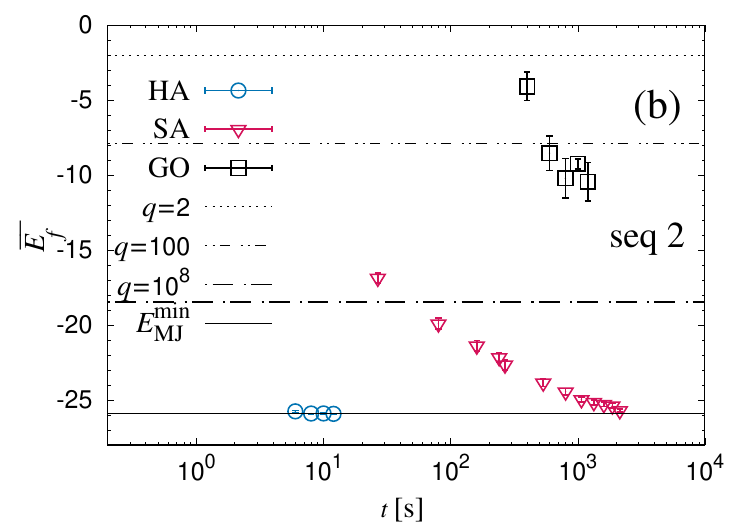}
   \includegraphics[width=5.4cm]{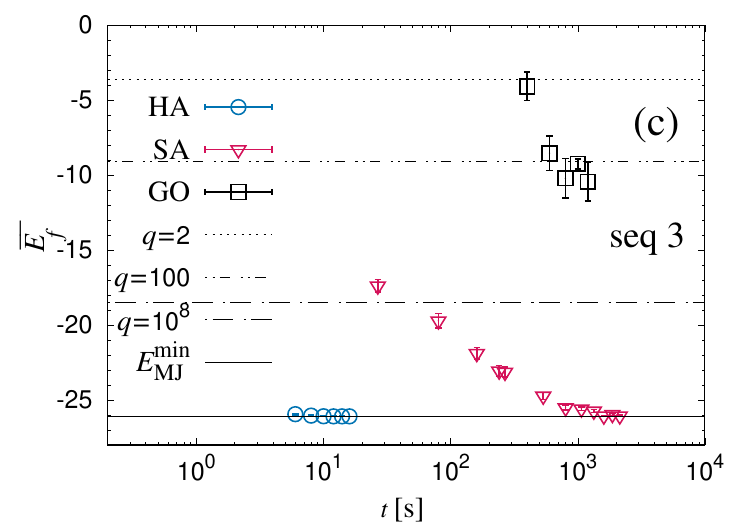}

   \includegraphics[width=5.4cm]{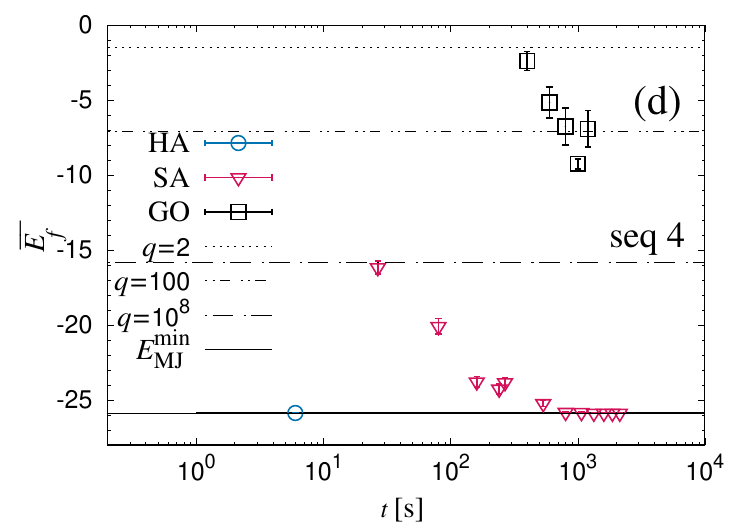}
   \includegraphics[width=5.4cm]{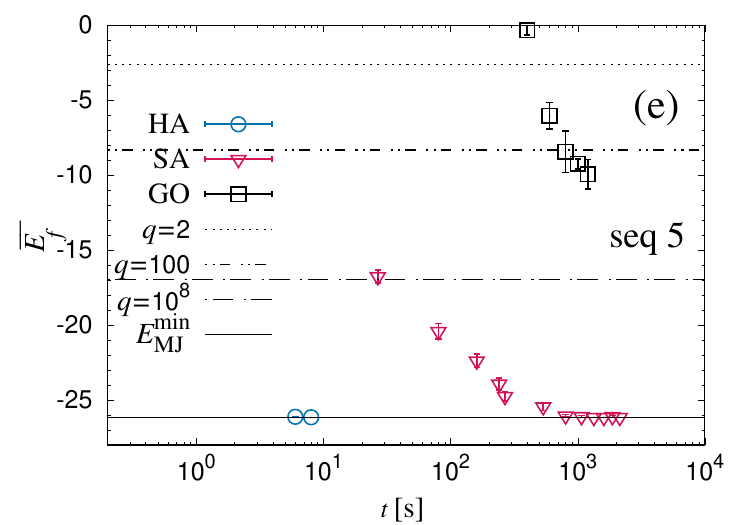}
   \includegraphics[width=5.4cm]{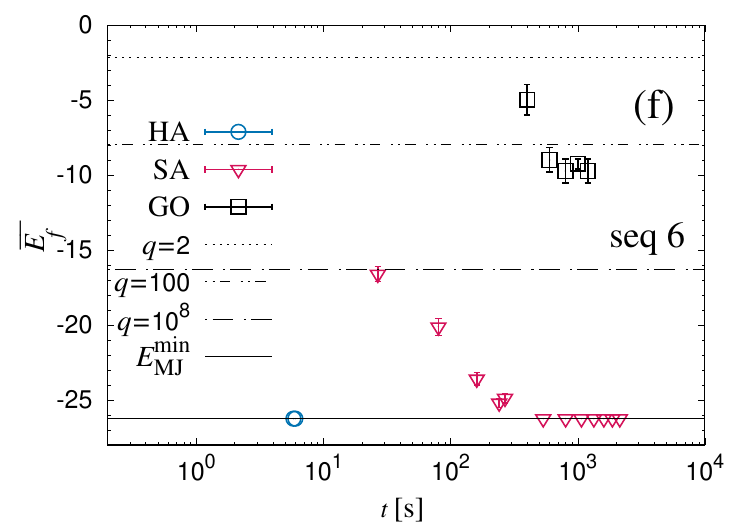}
\caption{Average final energy, $\ovEf$, plotted against run time, $t$, in HA, SA and GO
computations for the six sequences in Table~\ref{tab:seq}, with Lagrange parameters $\lv=(1.5,2.0,2.0)$ [Eq.~(\ref{eq:E})].
Each data point represent an average over 10 (GO) or 100 (HA and SA) independent runs. The horizontal lines indicate
data on the distribution of $\EMJ$ over valid chain structures, as obtained by exhaustive
enumerations (Sec.~\ref{sec:methods_enumeration}).
The lowest line represents $\EMJ^{\min}$. The other three represent
the lowest $q$-quantiles for $q=2$ (median), $q=100$ and $q=10^8$, respectively.
(a) Sequence 1. (b) Sequence 2. (c) Sequence 3. (d) Sequence 4. (e) Sequence 5. (f) Sequence 6.
\label{fig:erelax}}
\end{figure}

Figure~\ref{fig:erelax} shows the run-time dependence of the average final energy in our
HA, SA and GO computations. At the longest run times, $\ovEf$ is close or
equal to $\EMJ^{\min}$ and falls well below the lowest $10^8$-quantile of the $\EMJ$ distribution
for all six sequences with both HA and SA. In fact, with HA, the success rate is one for all six sequences.
With SA, the success rate is 1 for the sequences 4--6, while falling between 0.87--0.98 for
the sequences 1--3. Thus, the sought solutions can readily found with both HA and SA.
By contrast, our GO computations failed to find any correct solutions. Indeed, comparing
the classical GO and SA methods for a given run time, $\ovEf$ is much higher with GO than SA (Fig.~\ref{fig:erelax}).

It is worth noting that the HA success rate is $\ge$0.6 for all six sequences even when using the shortest run time possible
on D-Wave's system (6\,s). In previous work on 2D lattice proteins~\cite{Irback:22},
the run time had to be increased above a system-dependent threshold in order to attain a
satisfactory success rate. For the systems studied in the present paper, it seems that the shortest possible run time is
above any such threshold.

For a given sequence, starting with the default run time (6\,s), we increased the HA run time in steps of 2\,s
until all 100 runs returned the correct solution. With this procedure, we found that the longest required run time varied between 6\,s and 22\,s, depending on sequence.

In SA, we used a fixed set of run times for all sequences. The longest runs, corresponding
to 80,000 Monte Carlo sweeps per temperature, took $2.1\times 10^3$\,s using one thread on an Intel Core i9-13900K processor.
This run time gave a success rate of one (sequences 4--6), or slightly below one (sequences 1--3; success rates between 0.87--0.98).

In sharp contrast to the HA and SA results, we were unable to find low-energy structures with the GO method~(Fig.~\ref{fig:erelax}).
In fact, the success rate was zero for all sequences and run times used. This poor performance can be traced to the
chain connectivity constraint, encoded by the penalty energy $E_3$ in Eq.~(\ref{eq:E}), 
which, unlike the other two constraints, cannot be
expressed in terms of linear equalities or inequalities. If this constraint is removed, one has a multi-component lattice gas of amino acids,
rather than a protein chain. For this gas, we found that GO worked very well. Having seen this, we also tried including the $E_3$ term in
the GO objective function, instead of treating chain connectivity as a hard constraint. However, the results were similarly poor.
In either case, virtually all improvements of the objective function in the GO runs occurred in steps based on heuristics.
To be able to successfully use GO for this problem, it appears that a different binary representation would have to be devised.


\subsection{Comparing HA and SA data with exact results\label{sec:results_densofstates}}

The number of maximally compact chain conformations on a $4\times4\times3$ lattice (chain length $N=48$),
$\approx$\,$1.3\times10^{11}$, is known
from previous work~\cite{Pande:94,Kloczkowski:97}. Using exhaustive enumerations techniques
(Sec.~\ref{sec:methods_enumeration} and Appendix~\ref{sec:app_enumeration}), we extended this work to
obtain the energies of all structures and thereby the exact density of states, $g(\EMJ$), for the
sequences in Table~\ref{tab:seq}. These results provide us with a ground truth with which the HA and SA results
can be compared. Generating similar results for a larger chain length would be challenging due to the
large processor and storage requirements. Therefore,
the chain length $N=48$ is well suited for testing the power of the QUBO approach.

\begin{figure}
\centering
   \includegraphics[width=5.4cm]{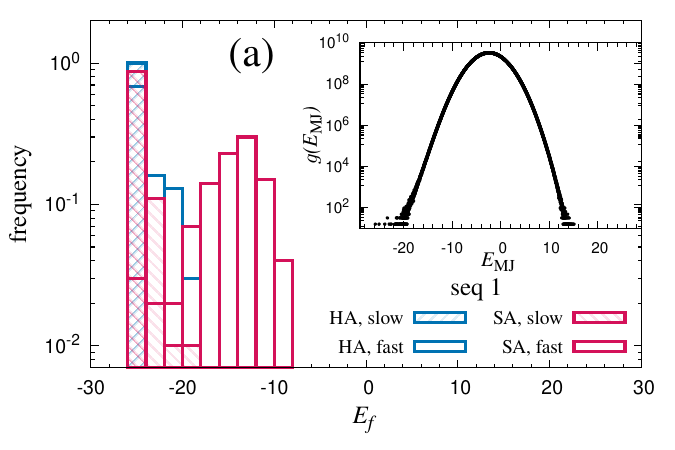}
   \includegraphics[width=5.4cm]{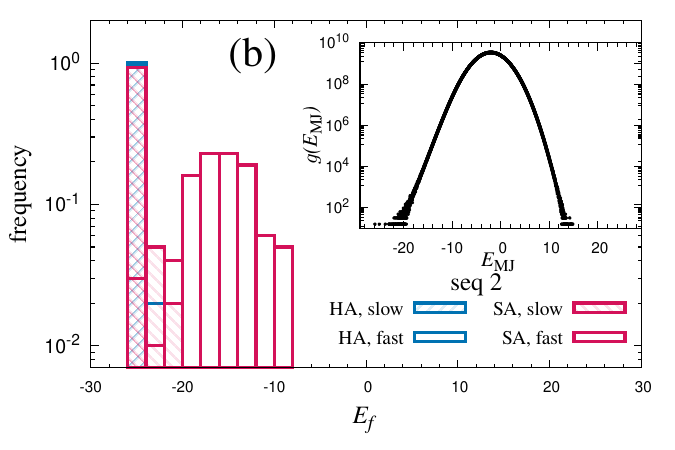}
   \includegraphics[width=5.4cm]{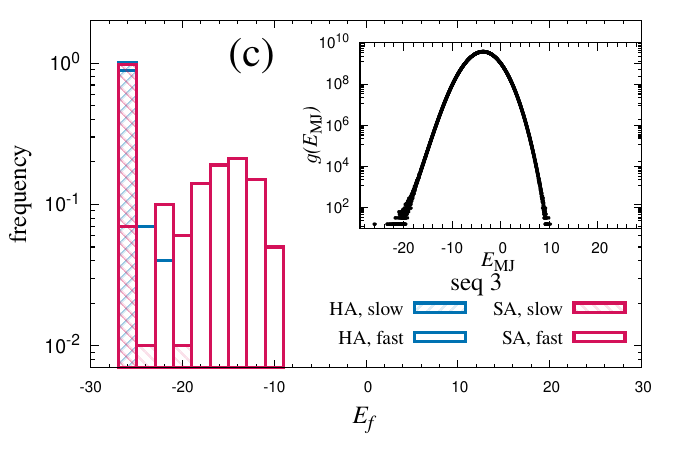}
   \includegraphics[width=5.4cm]{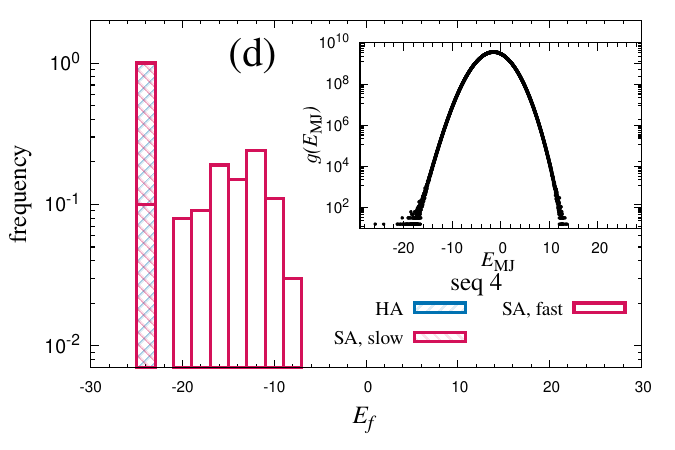}
   \includegraphics[width=5.4cm]{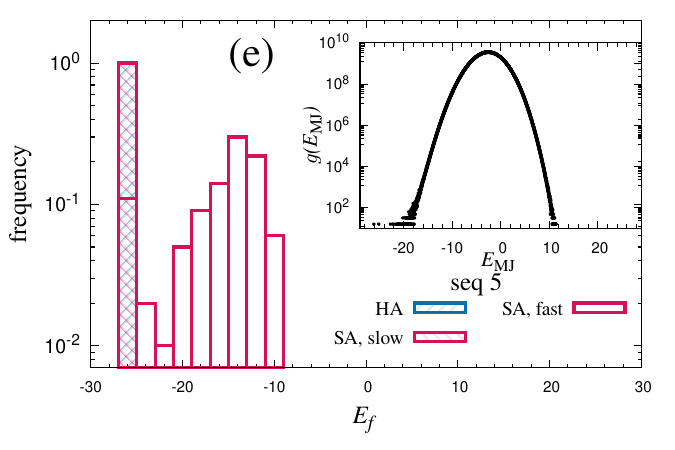}
   \includegraphics[width=5.4cm]{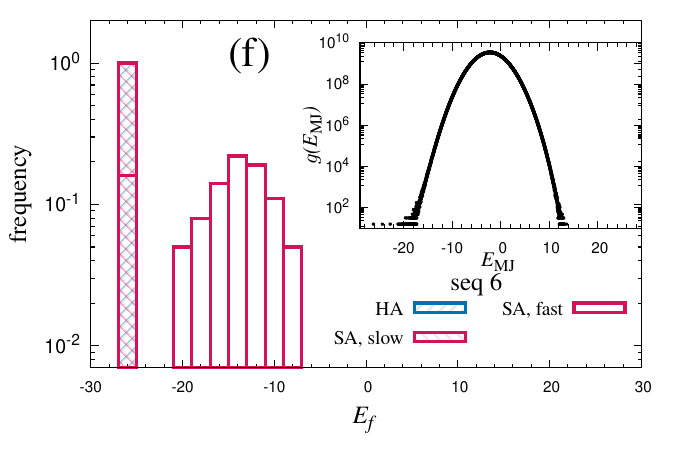}
\caption{Histograms of the final energy $E_f$ [Eq.~(\ref{eq:E})] on a log scale for HA and SA computations
for the six sequences in Table~\ref{tab:seq}, for the shortest (fast) and longest (slow) run times used.
 The insets show the density of states (log scale) calculated
as a function of $\EMJ$.
(a) Sequence 1. (b) Sequence 2. (c) Sequence 3. (d) Sequence 4.
(e) Sequence 5. (f) Sequence 6.
\label{fig:enum_and_dens}}
\end{figure}

Figure~\ref{fig:enum_and_dens} shows histograms of the final energy $E_f$ in the HA and SA runs
for our shortest and longest run times for the different sequences. Also shown, for comparison, are the
densities of states, $g(\EMJ$) (insets). The $g(\EMJ$) data has an effectively continuous and approximately
Gaussian central part where the bulk of the states are found, with an inflection point on the low-energy arm of the
curve (at $\EMJ\approx -17$). The states of main interest to us are the lowest-lying ones, which form a
discrete tail of the distribution.

The final energies $E_f$ of the HA runs are consistently located in the discrete low-energy tail of $g(\EMJ)$,
even for the shortest run time (6\,s). Increasing this annealing time to around 4 times the minimum allowed
value results in $\ovEf=\EMJ^{\min}$ for the HA runs. For SA with the shortest run time (27\,s), a significant fraction the runs
have final energies $E_f$ in the continuous part of the distribution. However, increasing the run time shifts the
$E_f$ distribution downward, as it should. At the longest run time ($2.1\times 10^3$\,s), most of the
runs end up in the desired minimum-$\EMJ$ state.

\begin{figure}
\centering
   \includegraphics[width=8cm]{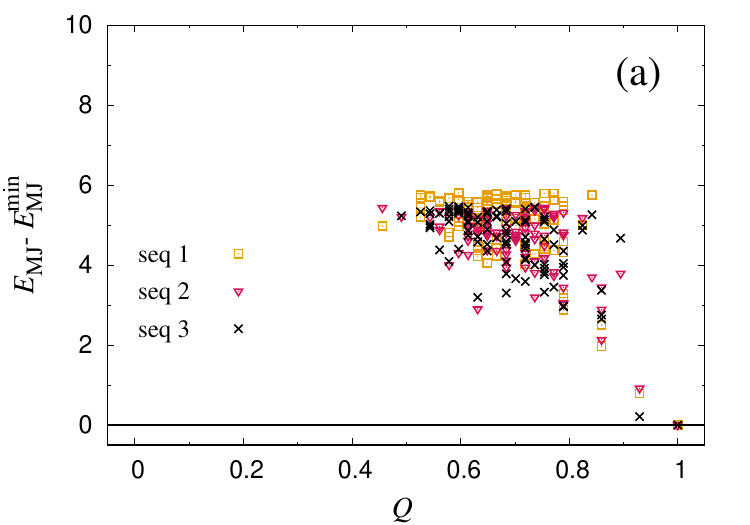}
   \includegraphics[width=8cm]{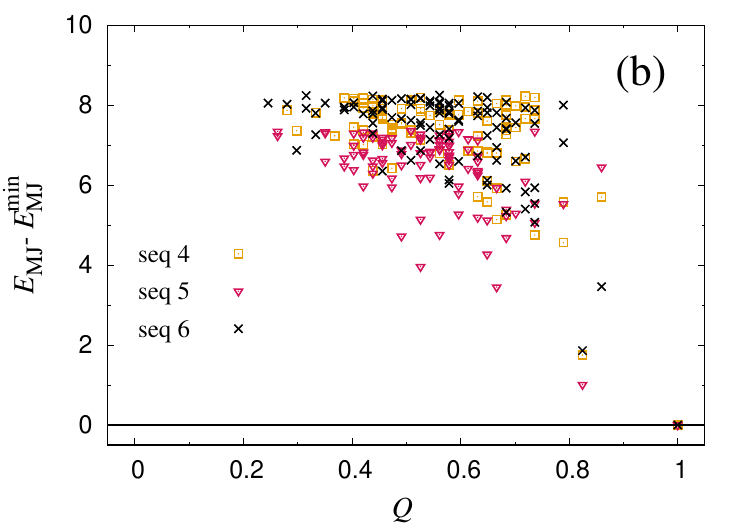}
\caption{Energy landscapes showing $\EMJ-\EMJ^{\min}$ against the overlap $Q$ with the native
state for the 100 lowest-lying maximally compact states for each of the six sequences in Table~\ref{tab:seq}, as obtained
through exhaustive enumeration (Sec.~\ref{sec:methods_enumeration}, Appendix~\ref{sec:app_enumeration}).
The overlap, or nativeness, $Q$ of a structure is the fraction of ground state contacts it contains. (a) Sequences 1--3, with low-complexity topology A as their native state.
(a) Sequences 4--6, with high-complexity topology B as their native state.
\label{fig:landscapes}}
\end{figure}

The ground state is given by topology A for three of the sequences (1--3) and by topology B for the other three
sequences (4--6). It is worth noting that the QUBO-based HA and SA methods solve the energy minimization
problem faster for the sequences associated with topology B than for those associated with topology A
(Fig.~\ref{fig:erelax}). By contrast, in explicit-chain Monte Carlo simulations of the same sequences on an
unrestricted grid, Fa\'\i sca and Plaxco found that folding occurred faster for the low-complexity topology A
than for the high-complexity topology B~\cite{Faisca:06}. To shed light on what causes this difference
between QUBO and explicit-chain results, we generated energy landscapes showing
energy against nativeness for the 100 lowest-lying maximally compact states (Fig.~\ref{fig:landscapes}). 
Here, the nativeness of a given structure is defined as the fraction of ground state contacts present in the structure.

Figure~\ref{fig:landscapes} shows the computed energy landscapes. While it is
important to remember that this analysis is restricted to maximally compact states, it is interesting 
that the sequences associated with the respective topologies A and B have 
notably different energy landscapes. For the former [Fig.~\ref{fig:landscapes}(a)], the data hints at a dense and funnel-like
organization of the low-lying states. This property may facilitate finding the native state with conventional explicit-chain
methods~\cite{Bryngelson:95} featuring diffusive state space exploration, which is consistent with 
the relatively fast folding of these sequences observed by Fa\'\i sca and Plaxco on an unrestricted grid~\cite{Faisca:06}. The high barriers between the
sparsely organized low-lying states of the high complexity systems 
[Fig.~\ref{fig:landscapes}(b)] pose a substantially harder challenge
for such approaches. With QUBO HA or SA the exploration of state space happens over an extended
conformation space containing additional unphysical states which seem to facilitate the search for
low-lying valid states for all our sequences. The clear separation in energy between the native and
competing states seems to facilitate the discovery of the ground states for sequences 4--6 in the
QUBO-based methods (Figs.~\ref{fig:erelax} and \ref{fig:landscapes}).


\section{Discussion\label{sec:discussion}}

Computational modeling of dense biomolecular systems is important to better understand the properties
of individual proteins under cellular conditions, and biophysical processes such as protein aggregation and
phase separation. A major obstacle in studying such problems with conventional explicit-chain
methods is steric clashes, which tend to make the exploration of state space very slow.
As a simple yet extreme example of a dense system, here we have considered the
task of determining minimum-energy structures of lattice proteins confined on a minimal grid, without any free sites.

The results presented show that this problem can be swiftly and consistently solved by recasting
it as a QUBO problem and using HA or SA. In fact, with HA, it was possible to find the correct solution
for chains with length $N=48$ in the order of 10\,s (Sec.~\ref{sec:results_timeevo}).

Traditionally, the problem of finding maximally compact lattice protein states with minimum energy
has been tackled through brute-force exhaustive enumerations. We implemented and used this
approach for our $4\times 4\times 3$ system. However, the calculations required a significant amount
of time (approximately 17\,h on 512 AMD EPYC 7742 processor cores) and storage, and would be cumbersome
to extend to longer chains due to the exponential growth of state space.

Another alternative is to stick to the binary description of the QUBO formulation, but replace the
soft constraints in favor of hard ones and use optimization methods based on linear/quadratic or constraint
programming. This approach was recently tested on two optimization problems somewhat related to ours,
namely lattice protein design~\cite{Panizza:24} and peptide docking~\cite{Brubaker:24}.
In both cases, these methods showed good results.
Therefore, we tested using the GO method for our problem. In stark contrast to the
good results obtained with this method for the design problem~\cite{Panizza:24}, GO failed
to find satisfactory solutions to our problem (Sec.~\ref{sec:results_timeevo}). To understand this
poor performance, we performed additional GO computations for systems where the chain connectivity
constraint, as encoded by the penalty energy $E_3$, was left out. For this system, corresponding
to a heterogeneous lattice gas, we found that GO worked very well. Hence, it appears that the poor
results obtained with GO for our lattice protein problem can be attributed to the presence of the chain
connectivity constraint. We note that this constraint, unlike the two encoded by the penalty energies $E_1$ and $E_2$,
cannot be expressed in terms of linear equalities or inequalities, and it is absent in the
design problem~\cite{Panizza:24}.

The folding rate of single-domain proteins is known to be inversely correlated with the contact order
of the native structure~\cite{Plaxco:98}. In line with this, previous work found folding to be faster for
our sequences 1--3 than for 4--6 (Table~\ref{tab:seq}) in kinetic explicit-chain Monte Carlo simulations
on an unrestricted grid~\cite{Faisca:06}. In our QUBO-based HA and SA computations, the opposite
trend is observed; the ground state is easier to locate for the sequences 4--6 than for 1--3 (Fig.~\ref{fig:erelax}).
This finding underscores that the extended QUBO systems need not follow intuition based on explicit-chain dynamics.

\section{Conclusion and outlook}

We have tested using a QUBO formulation, amenable to quantum optimization,
to solve the simplified yet computationally challenging problem of determining maximally compact
minimum-energy structures of lattice proteins. The results presented show that this problem can be
swiftly and successfully solved with the QUBO approach using either classical (SA) or hybrid quantum-classical
(HA) computations, for a chain length of 48. In fact, with HA, it took the order of 10\,s to obtain correct solutions.
We are not aware of any existing competing method with a potentially similar performance.

The QUBO-based computations presented here can be easily extended to dense multi-chain systems,
as in protein aggregation~\cite{Abeln:14} or phase separation~\cite{Rana:21}. In their present form,
the methods assume, however, a lattice-based biophysical model. Whether or not they
can be extended to continuous models remains to be seen.

The QUBO approach, which was employed for optimization
problems already in the 1980s~\cite{Hopfield:85,Peterson:88}, may be relevant for problems seemingly
different from the ones studied here. Since it opens up a non-physical extension to the state space,
it can be utilized to ``walk around" high energy barriers. In particular, other problems where constraint-fulfilling
updates are difficult, such as tight scheduling problems~\cite{Lagerholm:97,Venturelli:16,Stollenwerk:20},
might benefit from such an encoding.


\section{Data availability}

During this study, we have generated and stored a listing of
all structures, up to symmetries, for an $N=48$
chain on a $4\times 4\times 3$ lattice. Since this list
represents the entire set of possible structures, the
information is independent of the sequences we have studied.
The archive with the structures, of size about 6.1 TB, is
available from us upon request.

\clearpage
\begin{acknowledgments}

We thank Carsten Peterson for fruitful discussions.
We gratefully acknowledge the J\"ulich Supercomputing Centre
(https://www.fz-juelich.de/ias/jsc) for supporting this project by providing computing time
on the D-Wave Advantage\texttrademark{} System JUPSI through the J\"ulich UNified
Infrastructure for Quantum computing (JUNIQ).

\end{acknowledgments}

\appendix 
\section{Choice of Lagrange parameters \label{sec:app_lagrange}}

The QUBO energy, Eq.~(\ref{eq:E}), contains three Lagrange parameters,
$\bm{\lambda}=(\lambda_1,\lambda_2,\lambda_3)$, which need to be set.
We investigated the dependence of the success rate on these parameters
in both QUBO HA and QUBO SA through computations for sequence~5 (Table~\ref{tab:seq}).
Here, we varied one $\lambda_i$ at a time in steps of $0.25$, while keeping the other two fixed.
The highest success rates were obtained using $\bm{\lambda}^*=(1.5,2.0,2.0)$.

\begin{figure}[t]
\centering
   \includegraphics[width=8cm]{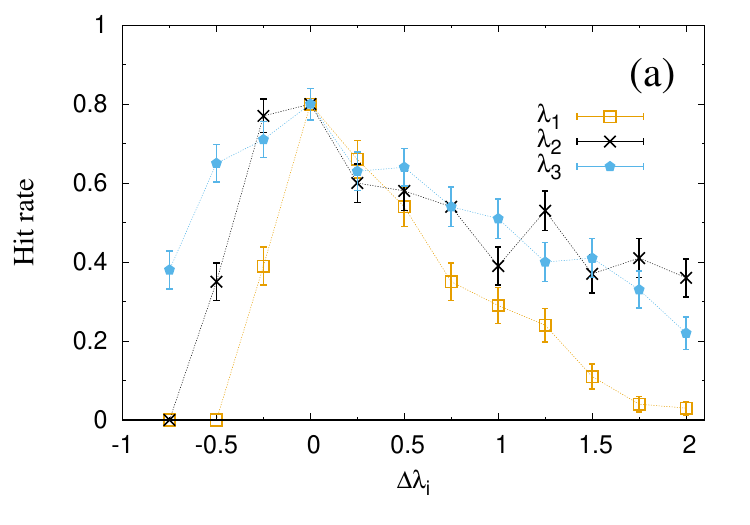}
   \includegraphics[width=8cm]{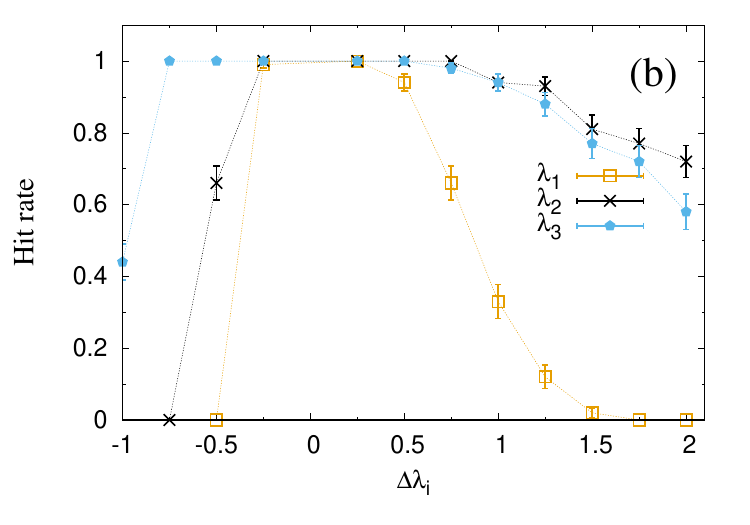}
   
\vspace{-12pt}   
\caption{Parameter dependence of the fraction of correct solutions (hit rate) in the vicinity of the
best Lagrange parameters found, $\bm{\lambda}^*=(1.5,2.0,2.0)$,
when using QUBO SA and QUBO HA to search for the ground state of sequence 5 (Table~1 in the main text). The hit rate is
plotted against $\Delta \lambda_i=\lambda_i-\lambda_i^*$, keeping $ \lambda_j= \lambda_j^*$
for $j\neq i$. Each data point represents an average over 100 runs. The SA runs comprised 10,000 sweeps per temperature, while
the HA run time was 10\,s. Lines are drawn to guide the eye. (a) SA (b) HA.
\label{fig:app_fig1}}
\end{figure}

Figure~\ref{fig:app_fig1} illustrates the sensitivity of the success rate to changes in
individual $\lambda_i$ parameters. Fortunately, the success rate is robust to small changes in
these parameters. Because of this robustness and the fact that our sequences share a
common length, we decided to use the same parameters for all six sequences, which worked well.

\section{Enumeration of all structures\label{sec:app_enumeration}}

We wish to enumerate all the $\approx$$1.3\times10^{11}$~\cite{Pande:94,Kloczkowski:97}
directed Hamiltonian paths unrelated by symmetry on a
$4\times 4\times 3$ lattice.
There is total of 16 symmetry operations on this lattice,
including the identity~\cite{Kloczkowski:97}. Let $(x,y,z)$ denote lattice coordinates,
where $x,y\in \{0,1,2,3\}$ and $z\in \{0,1,2\}$.

By symmetry, there are six possible starting points for the paths, which may be taken as
$(0,0,0)$, $(1,0,0)$, $(1,1,0)$, $(0,0,1)$, $(1,0,1)$ and $(1,1,1)$. After selecting starting point,
one or both of two symmetries may remain unbroken. One remaining symmetry operation is reflection
across the $(x,x,z)$ plane, which leaves the starting points $(0,0,0)$, $(0,0,1)$,
$(1,1,0)$ and $(1,1,1)$ unchanged. To break this symmetry, we note that when a path starting at
$(x_0,y_0,z_0)$ leaves the line $(x_0,y_0,z)$ for the first time, it does so in one of four
directions, $\pm\hat\mathbf{x}$ and  $\pm\hat\mathbf{y}$, where $+\hat\mathbf{x},+\hat\mathbf{y}$
and $-\hat\mathbf{x},-\hat\mathbf{y}$ are symmetry-related pairs. The symmetry is broken
by allowing only one direction from each pair, say  $+\hat\mathbf{x}$ and $-\hat\mathbf{y}$.
The other remaining symmetry operation is reflection across the $(x,y,1)$ plane, which leaves the starting points
$(0,0,1)$, $(0,0,1)$ and $(1,1,1)$ unchanged. This symmetry is broken by stipulating that the first
step away from the $(x, y, 1)$ plane is in, say, the $-\hat\mathbf{z}$ direction.

To generate the desired paths, we proceed in two steps. The first step is a width-first search,
where symmetries are accounted for by using the starting points and rules just described. Given
all self-avoiding paths with length $n-1$, we generate and store all with length $n$. This is iterated
up to path length $n=20$.  This is a very memory intensive approach. In the second step, we
therefore switch to a depth-first search to add the remaining 28 nodes to the paths. For each of the
seed paths obtained in the previous step, we recursively extend the structure until we exhaust all
lattice sites or until we reach a dead end. The heart of the procedure is a function to explore all
extensions of a partial path of length $n$. Given an input path of length $n$, the function loops
over all neighbors of the last node in the input path. If a neighbor appears anywhere in the path up to position $n-1$, 
it is discarded. Otherwise, we create a new path by adding it to the input path.
We then recursively call the same function to explore all extensions of the incremented path. Whenever
a path length $n=48$ is given as an input to the function, it stores it, and returns without any further
recursive calls. The simple loop and recursion algorithm above presents easy opportunities for
parallelisation. We used OneAPI Threading Building Blocks (onetbb) for parallelization of the
structure search.

Storage and post-processing of the very large number of structures in problems like this present
their own challenges. We devised an obvious compact representation for the explored paths, requiring
48 bytes per path. Despite this, the files with explored paths grew to many TB in size.
However, our recursive
procedure stores similar paths close to each other in the file, which allows for very large compression
factors to be used (for instance with \texttt{xz}) for long term storage.

%


\begin{thebibliography}{41}%
\makeatletter
\providecommand \@ifxundefined [1]{%
 \@ifx{#1\undefined}
}%
\providecommand \@ifnum [1]{%
 \ifnum #1\expandafter \@firstoftwo
 \else \expandafter \@secondoftwo
 \fi
}%
\providecommand \@ifx [1]{%
 \ifx #1\expandafter \@firstoftwo
 \else \expandafter \@secondoftwo
 \fi
}%
\providecommand \natexlab [1]{#1}%
\providecommand \enquote  [1]{``#1''}%
\providecommand \bibnamefont  [1]{#1}%
\providecommand \bibfnamefont [1]{#1}%
\providecommand \citenamefont [1]{#1}%
\providecommand \href@noop [0]{\@secondoftwo}%
\providecommand \href [0]{\begingroup \@sanitize@url \@href}%
\providecommand \@href[1]{\@@startlink{#1}\@@href}%
\providecommand \@@href[1]{\endgroup#1\@@endlink}%
\providecommand \@sanitize@url [0]{\catcode `\\12\catcode `\$12\catcode
  `\&12\catcode `\#12\catcode `\^12\catcode `\_12\catcode `\%12\relax}%
\providecommand \@@startlink[1]{}%
\providecommand \@@endlink[0]{}%
\providecommand \url  [0]{\begingroup\@sanitize@url \@url }%
\providecommand \@url [1]{\endgroup\@href {#1}{\urlprefix }}%
\providecommand \urlprefix  [0]{URL }%
\providecommand \Eprint [0]{\href }%
\providecommand \doibase [0]{https://doi.org/}%
\providecommand \selectlanguage [0]{\@gobble}%
\providecommand \bibinfo  [0]{\@secondoftwo}%
\providecommand \bibfield  [0]{\@secondoftwo}%
\providecommand \translation [1]{[#1]}%
\providecommand \BibitemOpen [0]{}%
\providecommand \bibitemStop [0]{}%
\providecommand \bibitemNoStop [0]{.\EOS\space}%
\providecommand \EOS [0]{\spacefactor3000\relax}%
\providecommand \BibitemShut  [1]{\csname bibitem#1\endcsname}%
\let\auto@bib@innerbib\@empty
\bibitem [{\citenamefont {Jumper}\ \emph {et~al.}(2021)\citenamefont {Jumper},
  \citenamefont {Evans}, \citenamefont {Pritzel}, \citenamefont {Green},
  \citenamefont {Figurnov}, \citenamefont {Ronneberger}, \citenamefont
  {Tunyasuvunakool}, \citenamefont {Bates}, \citenamefont {?'dek},
  \citenamefont {Potapenko}, \citenamefont {Bridgland}, \citenamefont {Meyer},
  \citenamefont {Kohl}, \citenamefont {Ballard}, \citenamefont {Cowie},
  \citenamefont {Romera-Paredes}, \citenamefont {Nikolov}, \citenamefont
  {Jain}, \citenamefont {Adler}, \citenamefont {Back}, \citenamefont
  {Petersen}, \citenamefont {Reiman}, \citenamefont {Clancy}, \citenamefont
  {Zielinski}, \citenamefont {Steinegger}, \citenamefont {Pacholska},
  \citenamefont {Berghammer}, \citenamefont {Bodenstein}, \citenamefont
  {Silver}, \citenamefont {Vinyals}, \citenamefont {Senior}, \citenamefont
  {Kavukcuoglu}, \citenamefont {Kohli},\ and\ \citenamefont
  {Hassabis}}]{Jumper:21}%
  \BibitemOpen
  \bibfield  {author} {\bibinfo {author} {\bibfnamefont {J.}~\bibnamefont
  {Jumper}}, \bibinfo {author} {\bibfnamefont {R.}~\bibnamefont {Evans}},
  \bibinfo {author} {\bibfnamefont {A.}~\bibnamefont {Pritzel}}, \bibinfo
  {author} {\bibfnamefont {T.}~\bibnamefont {Green}}, \bibinfo {author}
  {\bibfnamefont {M.}~\bibnamefont {Figurnov}}, \bibinfo {author}
  {\bibfnamefont {O.}~\bibnamefont {Ronneberger}}, \bibinfo {author}
  {\bibfnamefont {K.}~\bibnamefont {Tunyasuvunakool}}, \bibinfo {author}
  {\bibfnamefont {R.}~\bibnamefont {Bates}}, \bibinfo {author} {\bibfnamefont
  {A.}~\bibnamefont {Žídek}}, \bibinfo {author} {\bibfnamefont
  {A.}~\bibnamefont {Potapenko}}, \bibinfo {author} {\bibfnamefont
  {A.}~\bibnamefont {Bridgland}}, \bibinfo {author} {\bibfnamefont
  {C.}~\bibnamefont {Meyer}}, \bibinfo {author} {\bibfnamefont {S.~A.~A.}\
  \bibnamefont {Kohl}}, \bibinfo {author} {\bibfnamefont {A.~J.}\ \bibnamefont
  {Ballard}}, \bibinfo {author} {\bibfnamefont {A.}~\bibnamefont {Cowie}},
  \bibinfo {author} {\bibfnamefont {B.}~\bibnamefont {Romera-Paredes}},
  \bibinfo {author} {\bibfnamefont {S.}~\bibnamefont {Nikolov}}, \bibinfo
  {author} {\bibfnamefont {R.}~\bibnamefont {Jain}}, \bibinfo {author}
  {\bibfnamefont {J.}~\bibnamefont {Adler}}, \bibinfo {author} {\bibfnamefont
  {T.}~\bibnamefont {Back}}, \bibinfo {author} {\bibfnamefont {S.}~\bibnamefont
  {Petersen}}, \bibinfo {author} {\bibfnamefont {D.}~\bibnamefont {Reiman}},
  \bibinfo {author} {\bibfnamefont {E.}~\bibnamefont {Clancy}}, \bibinfo
  {author} {\bibfnamefont {M.}~\bibnamefont {Zielinski}}, \bibinfo {author}
  {\bibfnamefont {M.}~\bibnamefont {Steinegger}}, \bibinfo {author}
  {\bibfnamefont {M.}~\bibnamefont {Pacholska}}, \bibinfo {author}
  {\bibfnamefont {T.}~\bibnamefont {Berghammer}}, \bibinfo {author}
  {\bibfnamefont {S.}~\bibnamefont {Bodenstein}}, \bibinfo {author}
  {\bibfnamefont {D.}~\bibnamefont {Silver}}, \bibinfo {author} {\bibfnamefont
  {O.}~\bibnamefont {Vinyals}}, \bibinfo {author} {\bibfnamefont {A.~W.}\
  \bibnamefont {Senior}}, \bibinfo {author} {\bibfnamefont {K.}~\bibnamefont
  {Kavukcuoglu}}, \bibinfo {author} {\bibfnamefont {P.}~\bibnamefont {Kohli}},\
  and\ \bibinfo {author} {\bibfnamefont {D.}~\bibnamefont {Hassabis}},\
  }\bibfield  {title} {\bibinfo {title} {Highly accurate protein structure
  prediction with {A}lpha{F}old},\ }\href@noop {} {\bibfield  {journal}
  {\bibinfo  {journal} {Nature}\ }\textbf {\bibinfo {volume} {596}},\ \bibinfo
  {pages} {583} (\bibinfo {year} {2021})}\BibitemShut {NoStop}%
\bibitem [{\citenamefont {Lau}\ and\ \citenamefont {Dill}(1989)}]{Lau:89}%
  \BibitemOpen
  \bibfield  {author} {\bibinfo {author} {\bibfnamefont {K.~F.}\ \bibnamefont
  {Lau}}\ and\ \bibinfo {author} {\bibfnamefont {K.~A.}\ \bibnamefont {Dill}},\
  }\bibfield  {title} {\bibinfo {title} {A lattice statistical mechanics model
  of the conformational and sequence spaces of proteins},\ }\href@noop {}
  {\bibfield  {journal} {\bibinfo  {journal} {Macromolecules}\ }\textbf
  {\bibinfo {volume} {22}},\ \bibinfo {pages} {3986} (\bibinfo {year}
  {1989})}\BibitemShut {NoStop}%
\bibitem [{\citenamefont {?ali}\ \emph {et~al.}(1994)\citenamefont {?ali},
  \citenamefont {Shakhnovich},\ and\ \citenamefont {Karplus}}]{Sali:94}%
  \BibitemOpen
  \bibfield  {author} {\bibinfo {author} {\bibfnamefont {A.}~\bibnamefont
  {?ali}}, \bibinfo {author} {\bibfnamefont {E.}~\bibnamefont {Shakhnovich}},\
  and\ \bibinfo {author} {\bibfnamefont {M.}~\bibnamefont {Karplus}},\
  }\bibfield  {title} {\bibinfo {title} {Kinetics of protein folding: A lattice
  model study of the requirements for folding to the native state},\
  }\href@noop {} {\bibfield  {journal} {\bibinfo  {journal} {J.\ Mol.\ Biol.}\
  }\textbf {\bibinfo {volume} {235}},\ \bibinfo {pages} {1614} (\bibinfo {year}
  {1994})}\BibitemShut {NoStop}%
\bibitem [{\citenamefont {Socci}\ and\ \citenamefont
  {Onuchic}(1994)}]{Socci:94}%
  \BibitemOpen
  \bibfield  {author} {\bibinfo {author} {\bibfnamefont {N.~D.}\ \bibnamefont
  {Socci}}\ and\ \bibinfo {author} {\bibfnamefont {J.~N.}\ \bibnamefont
  {Onuchic}},\ }\bibfield  {title} {\bibinfo {title} {Folding kinetics of
  proteinlike heteropolymers},\ }\href@noop {} {\bibfield  {journal} {\bibinfo
  {journal} {J.\ Chem.\ Phys.}\ }\textbf {\bibinfo {volume} {101}},\ \bibinfo
  {pages} {1519} (\bibinfo {year} {1994})}\BibitemShut {NoStop}%
\bibitem [{\citenamefont {Klimov}\ and\ \citenamefont
  {Thirumalai}(1995)}]{Klimov:95}%
  \BibitemOpen
  \bibfield  {author} {\bibinfo {author} {\bibfnamefont {D.~K.}\ \bibnamefont
  {Klimov}}\ and\ \bibinfo {author} {\bibfnamefont {D.}~\bibnamefont
  {Thirumalai}},\ }\bibfield  {title} {\bibinfo {title} {Criterion that
  determines the foldability of proteins},\ }\href@noop {} {\bibfield
  {journal} {\bibinfo  {journal} {Phys.\ Rev.\ Lett.}\ }\textbf {\bibinfo
  {volume} {76}},\ \bibinfo {pages} {4070} (\bibinfo {year}
  {1995})}\BibitemShut {NoStop}%
\bibitem [{\citenamefont {Li}\ \emph {et~al.}(1996)\citenamefont {Li},
  \citenamefont {Helling}, \citenamefont {Tang},\ and\ \citenamefont
  {Wingreen}}]{Li:96}%
  \BibitemOpen
  \bibfield  {author} {\bibinfo {author} {\bibfnamefont {H.}~\bibnamefont
  {Li}}, \bibinfo {author} {\bibfnamefont {R.}~\bibnamefont {Helling}},
  \bibinfo {author} {\bibfnamefont {C.}~\bibnamefont {Tang}},\ and\ \bibinfo
  {author} {\bibfnamefont {N.}~\bibnamefont {Wingreen}},\ }\bibfield  {title}
  {\bibinfo {title} {Emergence of preferred structures in a simple model of
  protein folding},\ }\href@noop {} {\bibfield  {journal} {\bibinfo  {journal}
  {Science}\ }\textbf {\bibinfo {volume} {273}},\ \bibinfo {pages} {666}
  (\bibinfo {year} {1996})}\BibitemShut {NoStop}%
\bibitem [{\citenamefont {Perdomo-Ortiz}\ \emph {et~al.}(2012)\citenamefont
  {Perdomo-Ortiz}, \citenamefont {Dickson}, \citenamefont {Drew-Brook},
  \citenamefont {Rose},\ and\ \citenamefont {Aspuru-Guzik}}]{Perdomo-Ortiz:12}%
  \BibitemOpen
  \bibfield  {author} {\bibinfo {author} {\bibfnamefont {A.}~\bibnamefont
  {Perdomo-Ortiz}}, \bibinfo {author} {\bibfnamefont {N.}~\bibnamefont
  {Dickson}}, \bibinfo {author} {\bibfnamefont {M.}~\bibnamefont {Drew-Brook}},
  \bibinfo {author} {\bibfnamefont {G.}~\bibnamefont {Rose}},\ and\ \bibinfo
  {author} {\bibfnamefont {A.}~\bibnamefont {Aspuru-Guzik}},\ }\bibfield
  {title} {\bibinfo {title} {Finding low-energy conformations of lattice
  protein models by quantum annealing},\ }\href@noop {} {\bibfield  {journal}
  {\bibinfo  {journal} {Sci. Rep.}\ }\textbf {\bibinfo {volume} {2}},\ \bibinfo
  {pages} {248} (\bibinfo {year} {2012})}\BibitemShut {NoStop}%
\bibitem [{\citenamefont {Robert}\ \emph {et~al.}(2021)\citenamefont {Robert},
  \citenamefont {Barkoutsos}, \citenamefont {Woerner},\ and\ \citenamefont
  {Tavernelli}}]{Robert:21}%
  \BibitemOpen
  \bibfield  {author} {\bibinfo {author} {\bibfnamefont {A.}~\bibnamefont
  {Robert}}, \bibinfo {author} {\bibfnamefont {P.~K.}\ \bibnamefont
  {Barkoutsos}}, \bibinfo {author} {\bibfnamefont {S.}~\bibnamefont
  {Woerner}},\ and\ \bibinfo {author} {\bibfnamefont {I.}~\bibnamefont
  {Tavernelli}},\ }\bibfield  {title} {\bibinfo {title} {Resource-efficient
  quantum algorithm for protein folding},\ }\href@noop {} {\bibfield  {journal}
  {\bibinfo  {journal} {Npj\ Quantum\ Inf.}\ }\textbf {\bibinfo {volume} {7}},\
  \bibinfo {pages} {38} (\bibinfo {year} {2021})}\BibitemShut {NoStop}%
\bibitem [{\citenamefont {Irb\"ack}\ \emph {et~al.}(2022)\citenamefont
  {Irb\"ack}, \citenamefont {Knuthson}, \citenamefont {Mohanty},\ and\
  \citenamefont {Peterson}}]{Irback:22}%
  \BibitemOpen
  \bibfield  {author} {\bibinfo {author} {\bibfnamefont {A.}~\bibnamefont
  {Irb\"ack}}, \bibinfo {author} {\bibfnamefont {L.}~\bibnamefont {Knuthson}},
  \bibinfo {author} {\bibfnamefont {S.}~\bibnamefont {Mohanty}},\ and\ \bibinfo
  {author} {\bibfnamefont {C.}~\bibnamefont {Peterson}},\ }\bibfield  {title}
  {\bibinfo {title} {Folding lattice proteins with quantum annealing},\
  }\href@noop {} {\bibfield  {journal} {\bibinfo  {journal} {Phys.\ Rev.\
  Res.}\ }\textbf {\bibinfo {volume} {4}},\ \bibinfo {pages} {043013} (\bibinfo
  {year} {2022})}\BibitemShut {NoStop}%
\bibitem [{\citenamefont {Hopfield}\ and\ \citenamefont
  {Tank}(1985)}]{Hopfield:85}%
  \BibitemOpen
  \bibfield  {author} {\bibinfo {author} {\bibfnamefont {J.~J.}\ \bibnamefont
  {Hopfield}}\ and\ \bibinfo {author} {\bibfnamefont {D.~W.}\ \bibnamefont
  {Tank}},\ }\bibfield  {title} {\bibinfo {title} {Neural computation of
  decisions in optimization problems},\ }\href@noop {} {\bibfield  {journal}
  {\bibinfo  {journal} {Biol.\ Cybern.}\ }\textbf {\bibinfo {volume} {52}},\
  \bibinfo {pages} {141} (\bibinfo {year} {1985})}\BibitemShut {NoStop}%
\bibitem [{\citenamefont {Peterson}\ and\ \citenamefont
  {Anderson}(1988)}]{Peterson:88}%
  \BibitemOpen
  \bibfield  {author} {\bibinfo {author} {\bibfnamefont {C.}~\bibnamefont
  {Peterson}}\ and\ \bibinfo {author} {\bibfnamefont {J.~R.}\ \bibnamefont
  {Anderson}},\ }\bibfield  {title} {\bibinfo {title} {Neural networks and
  {NP}-complete problems; a performance study of the graph bisectioning
  problem},\ }\href@noop {} {\bibfield  {journal} {\bibinfo  {journal} {Complex
  Syst.}\ }\textbf {\bibinfo {volume} {2}},\ \bibinfo {pages} {59} (\bibinfo
  {year} {1988})}\BibitemShut {NoStop}%
\bibitem [{\citenamefont {Kadowaki}\ and\ \citenamefont
  {Nishimori}(1998)}]{Kadowaki:98}%
  \BibitemOpen
  \bibfield  {author} {\bibinfo {author} {\bibfnamefont {T.}~\bibnamefont
  {Kadowaki}}\ and\ \bibinfo {author} {\bibfnamefont {H.}~\bibnamefont
  {Nishimori}},\ }\bibfield  {title} {\bibinfo {title} {Quantum annealing in
  the transverse {I}sing model},\ }\href@noop {} {\bibfield  {journal}
  {\bibinfo  {journal} {Phys.\ Rev.\ E}\ }\textbf {\bibinfo {volume} {58}},\
  \bibinfo {pages} {5355} (\bibinfo {year} {1998})}\BibitemShut {NoStop}%
\bibitem [{\citenamefont {Farhi}\ \emph {et~al.}(2001)\citenamefont {Farhi},
  \citenamefont {Goldstone}, \citenamefont {Gutmann}, \citenamefont {Lapan},
  \citenamefont {Lundgren},\ and\ \citenamefont {Preda.}}]{Fahri:01}%
  \BibitemOpen
  \bibfield  {author} {\bibinfo {author} {\bibfnamefont {E.}~\bibnamefont
  {Farhi}}, \bibinfo {author} {\bibfnamefont {J.}~\bibnamefont {Goldstone}},
  \bibinfo {author} {\bibfnamefont {S.}~\bibnamefont {Gutmann}}, \bibinfo
  {author} {\bibfnamefont {J.}~\bibnamefont {Lapan}}, \bibinfo {author}
  {\bibfnamefont {A.}~\bibnamefont {Lundgren}},\ and\ \bibinfo {author}
  {\bibfnamefont {D.}~\bibnamefont {Preda.}},\ }\bibfield  {title} {\bibinfo
  {title} {A quantum adiabatic evolution algorithm applied to random instances
  of an np-complete problem},\ }\href@noop {} {\bibfield  {journal} {\bibinfo
  {journal} {Science}\ }\textbf {\bibinfo {volume} {292}},\ \bibinfo {pages}
  {472} (\bibinfo {year} {2001})}\BibitemShut {NoStop}%
\bibitem [{\citenamefont {Farhi}\ \emph {et~al.}(2014)\citenamefont {Farhi},
  \citenamefont {Goldstone},\ and\ \citenamefont {Gutmann}}]{Fahri:14}%
  \BibitemOpen
  \bibfield  {author} {\bibinfo {author} {\bibfnamefont {E.}~\bibnamefont
  {Farhi}}, \bibinfo {author} {\bibfnamefont {J.}~\bibnamefont {Goldstone}},\
  and\ \bibinfo {author} {\bibfnamefont {S.}~\bibnamefont {Gutmann}},\
  }\bibfield  {title} {\bibinfo {title} {A quantum approximate optimization
  algorithm},\ }\href@noop {} {\bibfield  {journal} {\bibinfo  {journal}
  {arXiv}\ ,\ \bibinfo {pages} {1411.4028}} (\bibinfo {year}
  {2014})}\BibitemShut {NoStop}%
\bibitem [{\citenamefont {Hadfield}\ \emph {et~al.}(2019)\citenamefont
  {Hadfield}, \citenamefont {Wang}, \citenamefont {OÕGorman}, \citenamefont
  {Rieffel}, \citenamefont {Venturelli},\ and\ \citenamefont
  {Biswas}}]{Hadfield:19}%
  \BibitemOpen
  \bibfield  {author} {\bibinfo {author} {\bibfnamefont {S.}~\bibnamefont
  {Hadfield}}, \bibinfo {author} {\bibfnamefont {Z.}~\bibnamefont {Wang}},
  \bibinfo {author} {\bibfnamefont {B.}~\bibnamefont {O'Gorman}}, \bibinfo
  {author} {\bibfnamefont {E.~G.}\ \bibnamefont {Rieffel}}, \bibinfo {author}
  {\bibfnamefont {D.}~\bibnamefont {Venturelli}},\ and\ \bibinfo {author}
  {\bibfnamefont {R.}~\bibnamefont {Biswas}},\ }\bibfield  {title} {\bibinfo
  {title} {From the quantum approximate optimization algorithm to a quantum
  alternating operator ansatz},\ }\href@noop {} {\bibfield  {journal} {\bibinfo
   {journal} {Algorithms}\ }\textbf {\bibinfo {volume} {12}},\ \bibinfo {pages}
  {34} (\bibinfo {year} {2019})}\BibitemShut {NoStop}%
\bibitem [{\citenamefont {Mohseni}\ \emph {et~al.}(2022)\citenamefont
  {Mohseni}, \citenamefont {McMahon},\ and\ \citenamefont
  {Byrnes}}]{Mohseni:22}%
  \BibitemOpen
  \bibfield  {author} {\bibinfo {author} {\bibfnamefont {N.}~\bibnamefont
  {Mohseni}}, \bibinfo {author} {\bibfnamefont {P.~L.}\ \bibnamefont
  {McMahon}},\ and\ \bibinfo {author} {\bibfnamefont {T.}~\bibnamefont
  {Byrnes}},\ }\bibfield  {title} {\bibinfo {title} {Ising machines as hardware
  solvers of combinatorial optimization problems},\ }\href@noop {} {\bibfield
  {journal} {\bibinfo  {journal} {Nat.\ Rev.\ Phys.}\ }\textbf {\bibinfo
  {volume} {4}},\ \bibinfo {pages} {363} (\bibinfo {year} {2022})}\BibitemShut
  {NoStop}%
\bibitem [{\citenamefont {Micheletti}\ \emph {et~al.}(2021)\citenamefont
  {Micheletti}, \citenamefont {Hauke},\ and\ \citenamefont
  {Faccioli}}]{Micheletti:21}%
  \BibitemOpen
  \bibfield  {author} {\bibinfo {author} {\bibfnamefont {C.}~\bibnamefont
  {Micheletti}}, \bibinfo {author} {\bibfnamefont {P.}~\bibnamefont {Hauke}},\
  and\ \bibinfo {author} {\bibfnamefont {P.}~\bibnamefont {Faccioli}},\
  }\bibfield  {title} {\bibinfo {title} {Polymer physics by quantum
  computing},\ }\href@noop {} {\bibfield  {journal} {\bibinfo  {journal}
  {Phys.\ Rev.\ Lett.}\ }\textbf {\bibinfo {volume} {127}},\ \bibinfo {pages}
  {080501} (\bibinfo {year} {2021})}\BibitemShut {NoStop}%
\bibitem [{\citenamefont {Slongo}\ \emph {et~al.}(2023)\citenamefont {Slongo},
  \citenamefont {Hauke}, \citenamefont {Faccioli},\ and\ \citenamefont
  {Micheletti}}]{Slongo:23}%
  \BibitemOpen
  \bibfield  {author} {\bibinfo {author} {\bibfnamefont {F.}~\bibnamefont
  {Slongo}}, \bibinfo {author} {\bibfnamefont {P.}~\bibnamefont {Hauke}},
  \bibinfo {author} {\bibfnamefont {P.}~\bibnamefont {Faccioli}},\ and\
  \bibinfo {author} {\bibfnamefont {C.}~\bibnamefont {Micheletti}},\ }\bibfield
   {title} {\bibinfo {title} {Quantum-inspired encoding enhances stochastic
  sampling of soft matter systems},\ }\href@noop {} {\bibfield  {journal}
  {\bibinfo  {journal} {Sci. Adv.}\ }\textbf {\bibinfo {volume} {9}},\ \bibinfo
  {pages} {eadi0204} (\bibinfo {year} {2023})}\BibitemShut {NoStop}%
\bibitem [{\citenamefont {Lagerholm}\ \emph {et~al.}(1997)\citenamefont
  {Lagerholm}, \citenamefont {Peterson},\ and\ \citenamefont
  {Sšderberg}}]{Lagerholm:97}%
  \BibitemOpen
  \bibfield  {author} {\bibinfo {author} {\bibfnamefont {M.}~\bibnamefont
  {Lagerholm}}, \bibinfo {author} {\bibfnamefont {C.}~\bibnamefont
  {Peterson}},\ and\ \bibinfo {author} {\bibfnamefont {B.}~\bibnamefont
  {S\"oderberg}},\ }\bibfield  {title} {\bibinfo {title} {Airline crew
  scheduling with {P}otts neurons},\ }\href@noop {} {\bibfield  {journal}
  {\bibinfo  {journal} {Neural\ Comput.}\ }\textbf {\bibinfo {volume} {9}},\
  \bibinfo {pages} {1589} (\bibinfo {year} {1997})}\BibitemShut {NoStop}%
\bibitem [{\citenamefont {Venturelli}\ \emph {et~al.}(2016)\citenamefont
  {Venturelli}, \citenamefont {Marchand},\ and\ \citenamefont
  {Rojo}}]{Venturelli:16}%
  \BibitemOpen
  \bibfield  {author} {\bibinfo {author} {\bibfnamefont {D.}~\bibnamefont
  {Venturelli}}, \bibinfo {author} {\bibfnamefont {D.~J.~J.}\ \bibnamefont
  {Marchand}},\ and\ \bibinfo {author} {\bibfnamefont {G.}~\bibnamefont
  {Rojo}},\ }\bibfield  {title} {\bibinfo {title} {Quantum annealing
  implementation of job-shop scheduling},\ }\href@noop {} {\bibfield  {journal}
  {\bibinfo  {journal} {arXiv}\ ,\ \bibinfo {pages} {1506.08479}} (\bibinfo
  {year} {2016})}\BibitemShut {NoStop}%
\bibitem [{\citenamefont {Stollenwerk}\ \emph {et~al.}(2020)\citenamefont
  {Stollenwerk}, \citenamefont {Hadfield},\ and\ \citenamefont
  {Wang}}]{Stollenwerk:20}%
  \BibitemOpen
  \bibfield  {author} {\bibinfo {author} {\bibfnamefont {T.}~\bibnamefont
  {Stollenwerk}}, \bibinfo {author} {\bibfnamefont {S.}~\bibnamefont
  {Hadfield}},\ and\ \bibinfo {author} {\bibfnamefont {Z.}~\bibnamefont
  {Wang}},\ }\bibfield  {title} {\bibinfo {title} {Toward quantum gate-model
  heuristics for real-world planning problems},\ }\href@noop {} {\bibfield
  {journal} {\bibinfo  {journal} {IEEE Trans. Quantum Eng.}\ }\textbf {\bibinfo
  {volume} {1}},\ \bibinfo {pages} {1} (\bibinfo {year} {2020})}\BibitemShut
  {NoStop}%
\bibitem [{\citenamefont {Kirkpatrick}\ \emph {et~al.}(1983)\citenamefont
  {Kirkpatrick}, \citenamefont {Gelatt},\ and\ \citenamefont
  {Vecchi}}]{Kirkpatrick:83}%
  \BibitemOpen
  \bibfield  {author} {\bibinfo {author} {\bibfnamefont {S.}~\bibnamefont
  {Kirkpatrick}}, \bibinfo {author} {\bibfnamefont {C.~D.}\ \bibnamefont
  {Gelatt}},\ and\ \bibinfo {author} {\bibfnamefont {M.~P.}\ \bibnamefont
  {Vecchi}},\ }\bibfield  {title} {\bibinfo {title} {Optimization by simulated
  annealing},\ }\href@noop {} {\bibfield  {journal} {\bibinfo  {journal}
  {Science}\ }\textbf {\bibinfo {volume} {220}},\ \bibinfo {pages} {671}
  (\bibinfo {year} {1983})}\BibitemShut {NoStop}%
\bibitem [{\citenamefont {McGeoch}\ \emph {et~al.}(2020)\citenamefont
  {McGeoch}, \citenamefont {Farr\'e},\ and\ \citenamefont
  {Bernoudy}}]{McGeoch:20b}%
  \BibitemOpen
  \bibfield  {author} {\bibinfo {author} {\bibfnamefont {C.}~\bibnamefont
  {McGeoch}}, \bibinfo {author} {\bibfnamefont {P.}~\bibnamefont {Farr\'e}},\
  and\ \bibinfo {author} {\bibfnamefont {W.}~\bibnamefont {Bernoudy}},\
  }\href@noop {} {\emph {\bibinfo {title} {D-Wave Hybrid Solver Service +
  Advantage: technology update}}},\ \bibinfo {type} {Tech. Rep.}\ (\bibinfo
  {institution} {D-Wave Systems Inc.},\ \bibinfo {year} {2020})\BibitemShut
  {NoStop}%
\bibitem [{\citenamefont {{Gurobi Optimization, LLC}}(2024)}]{gurobi}%
  \BibitemOpen
  \bibfield  {author} {\bibinfo {author} {\bibnamefont {{Gurobi Optimization,
  LLC}}},\ }\href@noop {} {\bibinfo {title} {{Gurobi Optimizer Reference
  Manual}}},\ \bibinfo {howpublished} {\url{https://www.gurobi.com/}} (\bibinfo
  {year} {2024})\BibitemShut {NoStop}%
\bibitem [{\citenamefont {Miyazawa}\ and\ \citenamefont
  {Jernigan}(1985)}]{Miyazawa:85}%
  \BibitemOpen
  \bibfield  {author} {\bibinfo {author} {\bibfnamefont {S.}~\bibnamefont
  {Miyazawa}}\ and\ \bibinfo {author} {\bibfnamefont {R.~L.}\ \bibnamefont
  {Jernigan}},\ }\bibfield  {title} {\bibinfo {title} {Estimation of effective
  interresidue contact energies from protein crystal structures: quasi-chemical
  approximation},\ }\href@noop {} {\bibfield  {journal} {\bibinfo  {journal}
  {Macromolecules}\ }\textbf {\bibinfo {volume} {18}},\ \bibinfo {pages} {534}
  (\bibinfo {year} {1985})}\BibitemShut {NoStop}%
\bibitem [{\citenamefont {Miyazawa}\ and\ \citenamefont
  {Jernigan}(1996)}]{Miyazawa:96}%
  \BibitemOpen
  \bibfield  {author} {\bibinfo {author} {\bibfnamefont {S.}~\bibnamefont
  {Miyazawa}}\ and\ \bibinfo {author} {\bibfnamefont {R.~L.}\ \bibnamefont
  {Jernigan}},\ }\bibfield  {title} {\bibinfo {title} {Residue-residue
  potentials with a favorable contact pair term and an unfavorable high packing
  density term, for simulation and threading},\ }\href@noop {} {\bibfield
  {journal} {\bibinfo  {journal} {J.\ Mol.\ Biol.}\ }\textbf {\bibinfo {volume}
  {256}},\ \bibinfo {pages} {623} (\bibinfo {year} {1996})}\BibitemShut
  {NoStop}%
\bibitem [{\citenamefont {Fa'sca}\ and\ \citenamefont
  {Plaxco}(2006)}]{Faisca:06}%
  \BibitemOpen
  \bibfield  {author} {\bibinfo {author} {\bibfnamefont {P.~F.}\ \bibnamefont
  {Faísca}}\ and\ \bibinfo {author} {\bibfnamefont {K.~W.}\ \bibnamefont
  {Plaxco}},\ }\bibfield  {title} {\bibinfo {title} {Cooperativity and the
  origins of rapid, single-exponential kinetics in protein folding},\
  }\href@noop {} {\bibfield  {journal} {\bibinfo  {journal} {Protein\ Sci.}\
  }\textbf {\bibinfo {volume} {15}},\ \bibinfo {pages} {1608} (\bibinfo {year}
  {2006})}\BibitemShut {NoStop}%
\bibitem [{\citenamefont {IrbŠck}\ \emph {et~al.}(2024)\citenamefont
  {IrbŠck}, \citenamefont {Knuthson}, \citenamefont {Mohanty},\ and\
  \citenamefont {Peterson}}]{Irback:24}%
  \BibitemOpen
  \bibfield  {author} {\bibinfo {author} {\bibfnamefont {A.}~\bibnamefont
  {Irb\"ack}}, \bibinfo {author} {\bibfnamefont {L.}~\bibnamefont {Knuthson}},
  \bibinfo {author} {\bibfnamefont {S.}~\bibnamefont {Mohanty}},\ and\ \bibinfo
  {author} {\bibfnamefont {C.}~\bibnamefont {Peterson}},\ }\bibfield  {title}
  {\bibinfo {title} {Using quantum annealing to design lattice proteins},\
  }\href@noop {} {\bibfield  {journal} {\bibinfo  {journal} {Phys.\ Rev.\
  Res.}\ }\textbf {\bibinfo {volume} {6}},\ \bibinfo {pages} {013162} (\bibinfo
  {year} {2024})}\BibitemShut {NoStop}%
\bibitem [{\citenamefont {Pande}\ \emph {et~al.}(1994)\citenamefont {Pande},
  \citenamefont {Joerg}, \citenamefont {Grosberg},\ and\ \citenamefont
  {Tanaka}}]{Pande:94}%
  \BibitemOpen
  \bibfield  {author} {\bibinfo {author} {\bibfnamefont {V.~S.}\ \bibnamefont
  {Pande}}, \bibinfo {author} {\bibfnamefont {C.}~\bibnamefont {Joerg}},
  \bibinfo {author} {\bibfnamefont {A.~Y.}\ \bibnamefont {Grosberg}},\ and\
  \bibinfo {author} {\bibfnamefont {T.}~\bibnamefont {Tanaka}},\ }\bibfield
  {title} {\bibinfo {title} {{Enumerations of the Hamiltonian walks on a cubic
  sublattice}},\ }\href@noop {} {\bibfield  {journal} {\bibinfo  {journal} {J.\
  Phys.\ A\ Math.\ Gen.}\ }\textbf {\bibinfo {volume} {27}},\ \bibinfo {pages}
  {6231} (\bibinfo {year} {1994})}\BibitemShut {NoStop}%
\bibitem [{\citenamefont {Kloczkowski}\ and\ \citenamefont
  {Jernigan}(1997)}]{Kloczkowski:97}%
  \BibitemOpen
  \bibfield  {author} {\bibinfo {author} {\bibfnamefont {A.}~\bibnamefont
  {Kloczkowski}}\ and\ \bibinfo {author} {\bibfnamefont {R.}~\bibnamefont
  {Jernigan}},\ }\bibfield  {title} {\bibinfo {title} {Computer generation and
  enumeration of compact self-avoiding walks within simple geometries on
  lattices},\ }\href@noop {} {\bibfield  {journal} {\bibinfo  {journal}
  {Comput.\ Theor.\ Polym.\ Sci.}\ }\textbf {\bibinfo {volume} {7}},\ \bibinfo
  {pages} {163} (\bibinfo {year} {1997})}\BibitemShut {NoStop}%
\bibitem [{\citenamefont {Plaxco}\ \emph {et~al.}(1998)\citenamefont {Plaxco},
  \citenamefont {Simons},\ and\ \citenamefont {Baker}}]{Plaxco:98}%
  \BibitemOpen
  \bibfield  {author} {\bibinfo {author} {\bibfnamefont {K.~W.}\ \bibnamefont
  {Plaxco}}, \bibinfo {author} {\bibfnamefont {K.~T.}\ \bibnamefont {Simons}},\
  and\ \bibinfo {author} {\bibfnamefont {D.}~\bibnamefont {Baker}},\ }\bibfield
   {title} {\bibinfo {title} {Contact order, transition state placement and the
  refolding rates of single domain proteins},\ }\href@noop {} {\bibfield
  {journal} {\bibinfo  {journal} {J.\ Mol.\ Biol.}\ }\textbf {\bibinfo {volume}
  {277}},\ \bibinfo {pages} {985} (\bibinfo {year} {1998})}\BibitemShut
  {NoStop}%
\bibitem [{\citenamefont {Linn}\ \emph {et~al.}(2024)\citenamefont {Linn},
  \citenamefont {Brundin}, \citenamefont {García-Álvarez},\ and\
  \citenamefont {Johansson}}]{Linn:24}%
  \BibitemOpen
  \bibfield  {author} {\bibinfo {author} {\bibfnamefont {H.}~\bibnamefont
  {Linn}}, \bibinfo {author} {\bibfnamefont {I.}~\bibnamefont {Brundin}},
  \bibinfo {author} {\bibfnamefont {L.}~\bibnamefont {García-Álvarez}},\ and\
  \bibinfo {author} {\bibfnamefont {G.}~\bibnamefont {Johansson}},\ }\bibfield
  {title} {\bibinfo {title} {{Resource analysis of quantum algorithms for
  coarse-grained protein folding models}},\ }\href@noop {} {\bibfield
  {journal} {\bibinfo  {journal} {Phys.\ Rev.\ Res.}\ }\textbf {\bibinfo
  {volume} {6}},\ \bibinfo {pages} {033112} (\bibinfo {year}
  {2024})}\BibitemShut {NoStop}%
\bibitem [{\citenamefont {Panizza}\ \emph {et~al.}(2025)\citenamefont
  {Panizza}, \citenamefont {Roggero}, \citenamefont {Hauke},\ and\
  \citenamefont {Faccioli}}]{Panizza:25}%
  \BibitemOpen
  \bibfield  {author} {\bibinfo {author} {\bibfnamefont {V.}~\bibnamefont
  {Panizza}}, \bibinfo {author} {\bibfnamefont {A.}~\bibnamefont {Roggero}},
  \bibinfo {author} {\bibfnamefont {P.}~\bibnamefont {Hauke}},\ and\ \bibinfo
  {author} {\bibfnamefont {P.}~\bibnamefont {Faccioli}},\ }\bibfield  {title}
  {\bibinfo {title} {Statistical mechanics of heteropolymers from lattice gauge
  theory},\ }\href@noop {} {\bibfield  {journal} {\bibinfo  {journal} {Phys.\
  Rev.\ Lett.}\ }\textbf {\bibinfo {volume} {134}},\ \bibinfo {pages} {158101}
  (\bibinfo {year} {2025})}\BibitemShut {NoStop}%
\bibitem [{\citenamefont {Dantzig}(1963)}]{Dantzig:63}%
  \BibitemOpen
  \bibfield  {author} {\bibinfo {author} {\bibfnamefont {G.~B.}\ \bibnamefont
  {Dantzig}},\ }\href {https://doi.org/doi:10.1515/9781400884179} {\emph
  {\bibinfo {title} {Linear Programming and Extensions}}}\ (\bibinfo
  {publisher} {Princeton University Press},\ \bibinfo {address} {Princeton},\
  \bibinfo {year} {1963})\BibitemShut {NoStop}%
\bibitem [{\citenamefont {Morrison}\ \emph {et~al.}(2016)\citenamefont
  {Morrison}, \citenamefont {Jacobson}, \citenamefont {Sauppe},\ and\
  \citenamefont {Sewell}}]{Morrison:16}%
  \BibitemOpen
  \bibfield  {author} {\bibinfo {author} {\bibfnamefont {D.~R.}\ \bibnamefont
  {Morrison}}, \bibinfo {author} {\bibfnamefont {S.~H.}\ \bibnamefont
  {Jacobson}}, \bibinfo {author} {\bibfnamefont {J.~J.}\ \bibnamefont
  {Sauppe}},\ and\ \bibinfo {author} {\bibfnamefont {E.~C.}\ \bibnamefont
  {Sewell}},\ }\bibfield  {title} {\bibinfo {title} {Branch-and-bound
  algorithms: A survey of recent advances in searching, branching, and
  pruning},\ }\href@noop {} {\bibfield  {journal} {\bibinfo  {journal}
  {Discrete\ Optim.}\ }\textbf {\bibinfo {volume} {19}},\ \bibinfo {pages} {79}
  (\bibinfo {year} {2016})}\BibitemShut {NoStop}%
\bibitem [{\citenamefont {Marchand}\ \emph {et~al.}(2002)\citenamefont
  {Marchand}, \citenamefont {Martin}, \citenamefont {Weismantel},\ and\
  \citenamefont {Wolsey}}]{Marchand:02}%
  \BibitemOpen
  \bibfield  {author} {\bibinfo {author} {\bibfnamefont {H.}~\bibnamefont
  {Marchand}}, \bibinfo {author} {\bibfnamefont {A.}~\bibnamefont {Martin}},
  \bibinfo {author} {\bibfnamefont {R.}~\bibnamefont {Weismantel}},\ and\
  \bibinfo {author} {\bibfnamefont {L.}~\bibnamefont {Wolsey}},\ }\bibfield
  {title} {\bibinfo {title} {Cutting planes in integer and mixed integer
  programming},\ }\href@noop {} {\bibfield  {journal} {\bibinfo  {journal}
  {Discrete\ Appl.\ Math.}\ }\textbf {\bibinfo {volume} {123}},\ \bibinfo
  {pages} {397} (\bibinfo {year} {2002})}\BibitemShut {NoStop}%
\bibitem [{\citenamefont {Panizza}\ \emph {et~al.}(2024)\citenamefont
  {Panizza}, \citenamefont {Hauke}, \citenamefont {Micheletti},\ and\
  \citenamefont {Faccioli}}]{Panizza:24}%
  \BibitemOpen
  \bibfield  {author} {\bibinfo {author} {\bibfnamefont {V.}~\bibnamefont
  {Panizza}}, \bibinfo {author} {\bibfnamefont {P.}~\bibnamefont {Hauke}},
  \bibinfo {author} {\bibfnamefont {C.}~\bibnamefont {Micheletti}},\ and\
  \bibinfo {author} {\bibfnamefont {P.}~\bibnamefont {Faccioli}},\ }\bibfield
  {title} {\bibinfo {title} {Protein design by integrating machine learning and
  quantum-encoded optimization},\ }\href@noop {} {\bibfield  {journal}
  {\bibinfo  {journal} {PRX Life}\ }\textbf {\bibinfo {volume} {2}},\ \bibinfo
  {pages} {043012} (\bibinfo {year} {2024})}\BibitemShut {NoStop}%
\bibitem [{\citenamefont {Bryngelson}\ \emph {et~al.}(1995)\citenamefont
  {Bryngelson}, \citenamefont {Onuchic}, \citenamefont {Socci},\ and\
  \citenamefont {Wolynes}}]{Bryngelson:95}%
  \BibitemOpen
  \bibfield  {author} {\bibinfo {author} {\bibfnamefont {J.~D.}\ \bibnamefont
  {Bryngelson}}, \bibinfo {author} {\bibfnamefont {J.~N.}\ \bibnamefont
  {Onuchic}}, \bibinfo {author} {\bibfnamefont {N.~D.}\ \bibnamefont {Socci}},\
  and\ \bibinfo {author} {\bibfnamefont {P.~G.}\ \bibnamefont {Wolynes}},\
  }\bibfield  {title} {\bibinfo {title} {Funnels, pathways, and the energy
  landscape of protein folding: A synthesis},\ }\href@noop {} {\bibfield
  {journal} {\bibinfo  {journal} {Proteins}\ }\textbf {\bibinfo {volume}
  {21}},\ \bibinfo {pages} {167} (\bibinfo {year} {1995})}\BibitemShut
  {NoStop}%
\bibitem [{\citenamefont {Brubaker}\ \emph {et~al.}(2024)\citenamefont
  {Brubaker}, \citenamefont {Booth}, \citenamefont {Arakawa}, \citenamefont
  {Furrer}, \citenamefont {Ghosh}, \citenamefont {Sato},\ and\ \citenamefont
  {Katzgraber}}]{Brubaker:24}%
  \BibitemOpen
  \bibfield  {author} {\bibinfo {author} {\bibfnamefont {J.~K.}\ \bibnamefont
  {Brubaker}}, \bibinfo {author} {\bibfnamefont {K.~E.~C.}\ \bibnamefont
  {Booth}}, \bibinfo {author} {\bibfnamefont {A.}~\bibnamefont {Arakawa}},
  \bibinfo {author} {\bibfnamefont {F.}~\bibnamefont {Furrer}}, \bibinfo
  {author} {\bibfnamefont {J.}~\bibnamefont {Ghosh}}, \bibinfo {author}
  {\bibfnamefont {T.}~\bibnamefont {Sato}},\ and\ \bibinfo {author}
  {\bibfnamefont {H.~G.}\ \bibnamefont {Katzgraber}},\ }\bibfield  {title}
  {\bibinfo {title} {Quadratic unconstrained binary optimization and constraint
  programming approaches for lattice-based cyclic peptide docking},\
  }\href@noop {} {\bibfield  {journal} {\bibinfo  {journal} {arXiv}\ }
  (\bibinfo {year} {2024})},\ \Eprint {https://arxiv.org/abs/2412.10260}
  {2412.10260} \BibitemShut {NoStop}%
\bibitem [{\citenamefont {Abeln}\ \emph {et~al.}(2014)\citenamefont {Abeln},
  \citenamefont {Vendruscolo}, \citenamefont {Dobson},\ and\ \citenamefont
  {Frenkel}}]{Abeln:14}%
  \BibitemOpen
  \bibfield  {author} {\bibinfo {author} {\bibfnamefont {S.}~\bibnamefont
  {Abeln}}, \bibinfo {author} {\bibfnamefont {M.}~\bibnamefont {Vendruscolo}},
  \bibinfo {author} {\bibfnamefont {C.~M.}\ \bibnamefont {Dobson}},\ and\
  \bibinfo {author} {\bibfnamefont {D.}~\bibnamefont {Frenkel}},\ }\bibfield
  {title} {\bibinfo {title} {A simple lattice model that captures protein
  folding, aggregation and amyloid formation},\ }\href@noop {} {\bibfield
  {journal} {\bibinfo  {journal} {PLoS\ One}\ }\textbf {\bibinfo {volume}
  {9}},\ \bibinfo {pages} {e85185} (\bibinfo {year} {2014})}\BibitemShut
  {NoStop}%
\bibitem [{\citenamefont {Rana}\ \emph {et~al.}(2021)\citenamefont {Rana},
  \citenamefont {Brangwynne},\ and\ \citenamefont {Panagiotopoulos}}]{Rana:21}%
  \BibitemOpen
  \bibfield  {author} {\bibinfo {author} {\bibfnamefont {U.}~\bibnamefont
  {Rana}}, \bibinfo {author} {\bibfnamefont {C.~P.}\ \bibnamefont
  {Brangwynne}},\ and\ \bibinfo {author} {\bibfnamefont {A.~Z.}\ \bibnamefont
  {Panagiotopoulos}},\ }\bibfield  {title} {\bibinfo {title} {Phase separation
  vs aggregation behavior for model disordered proteins},\ }\href@noop {}
  {\bibfield  {journal} {\bibinfo  {journal} {J.\ Chem.\ Phys.}\ }\textbf
  {\bibinfo {volume} {155}},\ \bibinfo {pages} {125101} (\bibinfo {year}
  {2021})}\BibitemShut {NoStop}%
\end{thebibliography}

\end{document}